\pdfoutput=1
\documentclass[iop,apj]{emulateapj}
\usepackage{color}
\usepackage{amsmath}
\usepackage{amssymb,bm}
\usepackage{xspace}
\usepackage[nolist,nohyperlinks]{acronym}
\usepackage{enumerate}
\usepackage{graphicx}
\usepackage{natbib}
\usepackage{multirow}
\usepackage{booktabs}
\usepackage[normalem]{ulem}

\newcommand{\mytilde}{\raise.17ex\hbox{$\scriptstyle\mathtt{\sim}$}}
\newcommand{\Msun}{\mbox{M$_\odot$}\xspace}

\newcommand{\dominik}{\citet{Dominik:2012kk}\xspace}

\newcommand{\change}[1]{{\color{black}#1}}
\renewcommand{\Re}{\operatorname{Re}}

\begin{document}
\title{Distinguishing compact binary population synthesis models using gravitational wave observations of coalescing binary black holes}


\author{Simon Stevenson}
\email{simon.stevenson@ligo.org}
\email{LIGO-P1500049}
\affil{School of Physics and Astronomy, University of Birmingham, Edgbaston,
Birmingham, B15 2TT}
\affil{School of Physics and Astronomy, Cardiff University, The Parade, Cardiff,
 CF24 3AA, UK}

\author{Frank Ohme}
\author{Stephen Fairhurst}
\affil{School of Physics and Astronomy, Cardiff University, The Parade, Cardiff,
 CF24 3AA, UK}

\begin{abstract}
The coalescence of compact binaries \change{containing} neutron stars or black holes is
one of the most promising signals for advanced ground-based laser interferometer
gravitational-wave detectors, with the first direct detections expected over the
next few years. The rate of binary coalescences and the distribution of
component masses is highly uncertain, and population synthesis models predict a
wide range of plausible values. Poorly constrained parameters in population
synthesis models correspond to poorly understood astrophysics at various stages
in the evolution of massive binary stars, the progenitors of binary neutron star
and binary black hole systems. These include effects such as supernova kick
velocities, parameters governing the energetics of common envelope evolution and
the strength of stellar winds.  Observing multiple binary black hole systems
through gravitational waves will allow us to infer details of the astrophysical
mechanisms that lead to their formation. Here we simulate gravitational-wave
observations from a series of population synthesis models including the effects
of known selection biases, measurement errors and cosmology. We compare the
predictions arising from different models and show that we will be able to distinguish between them with observations (or
the lack of them) from the early runs of the advanced LIGO and Virgo detectors.
This will allow us to narrow down the large parameter space for binary evolution
models.
\end{abstract}


\maketitle

\acrodef{PN}{post-Newtonian}
\acrodef{GW}{gravitational-wave}
\acrodef{BBH}{binary black hole}
\acrodef{BNS}{binary neutron star}
\acrodef{NSBH}{neutron star-black hole}
\acrodef{SNR}{signal-to-noise ratio}
\acrodef{IMF}{initial-mass-function}
\acrodef{aLIGO}{Advanced LIGO}
\acrodef{AdV}{Advanced Virgo}
\acrodef{MWEG}{Milky Way Equivalent Galaxy}
\acrodef{CE}{Common-envelope}

\newcommand{\PN}[0]{\ac{PN}\xspace}
\newcommand{\BBH}[0]{\ac{BBH}\xspace}
\newcommand{\BNS}[0]{\ac{BNS}\xspace}
\newcommand{\NSBH}[0]{\ac{NSBH}\xspace}
\newcommand{\GW}[0]{\ac{GW}\xspace}
\newcommand{\SNR}[0]{\ac{SNR}\xspace}
\newcommand{\IMF}[0]{\ac{IMF}\xspace}
\newcommand{\aLIGO}[0]{\ac{aLIGO}\xspace}
\newcommand{\AdV}[0]{\ac{AdV}\xspace}
\newcommand{\MWEG}[0]{\ac{MWEG}\xspace}
\newcommand{\CE}[0]{\ac{CE}\xpsace}

\section{Introduction}

The \aLIGO \citep{2015CQGra..32g4001T} 
and \AdV \citep{TheVirgo:2014hva} second generation,
kilometre-scale
ground based laser interferometers are currently being commissioned and
should begin observing runs in 2015 \citep{Aasi:2013wya} with the sensitivity
increasing gradually over a number of years before reaching their design
sensitivity near the end of the decade.  These \GW
observatories will be an order of magnitude more sensitive than the first
generation observatories and are expected to yield the first \GW detections 
\citep{Abadie:2010cf}
and herald the beginning of \GW astronomy. In \GW astronomy we are interested in the emission of
gravitational radiation from astrophysical sources. One of the primary sources
of \acp{GW} for \aLIGO is the coalescence of compact binaries --
\BNS, \NSBH and \BBH systems. 

The orbits of these systems decay due to radiation reaction
\citep{PhysRev.131.435, Peters:1964}, causing the two objects to
spiral in towards one another. During the final orbits and merger, 
these sources emit a large amount of gravitational radiation, and
this will be observable by \aLIGO and \AdV.
The gravitational waveform emitted by the binary can be modelled with great accuracy using the post-Newtonian formalism
\citep{lrr-2006-4}.  Closer to merger, full numerical simulations are required to
track the binary evolution and calculate the waveform (see 
\citet{Hannam:2009rd,Hinder:2010vn,Sperhake:2011xk} for overviews).  
By combining the insights of post-Newtonian theory and numerical modelling,
a large range of analytic/semi-analytic approximate waveform models
have been developed over the past few years (see, e.g., \citet{Buonanno:2009zt} and
\citet{Ohme:2011rm} for an overview). These models now provide accurate waveforms 
over a large fraction of the parameter space of non-precessing \acp{BBH}.
In particular, they provide accurate waveforms for signals with a range of mass ratios and
also cover the space of aligned spins.  There is ongoing work
\citep{Hannam:2013oca,Pan:2013rra} 
to extend these
to the full parameter space that incorporates spin-induced precession of the binary orbit.

The availability of accurate waveform models makes a matched filter search of these
signals feasible \citep{Babak:2012zx, Aasi:2012rja} and allows us to to extract
the physical
parameters of the binary system from the observed \GW signal 
\citep{Aasi:2013jjl,Veitch:2014wba}.  The observed sky location and orientation of the binary
system will be used to aid searches for electromagnetic counterparts of \GW systems 
\citep{Virgo:2011aa, Singer:2014qca, Aasi:2014iia, Clark:2014ut}.  Meanwhile,
measurement of the masses and spins of the binary components will shed light
upon the formation and evolution of the binary by comparing the observations
with predictions from stellar evolution models.
We expect the majority of systems to be observed with relatively low \SNR and consequently the parameters will not be measured with great
accuracy \citep{Hannam:2013uu,Ohme:2013nsa}. For an individual binary,
the \textit{chirp mass} of the system --- a combination of the two masses that
determines the rate at which the binary evolves --- can be measured with good accuracy
\citep{Cutler:1994ys, Hannam:2013uu}, while the mass ratio and spins are unlikely to be well
constrained. 

In addition, there is significant uncertainty in the astrophysical mass and spin distributions of black hole binaries.  Thus, it seems unlikely that the measurement of parameters from individual systems will significantly impact our understanding of black hole binary formation.  Instead, it will require the
measurement of parameters from a population of signals to significantly constrain
compact binary formation and evolution models.  In this paper, we consider how this
might be done and what we expect to learn with the observations from the early
\aLIGO and \AdV runs.

Compact binaries can be formed as a result of the evolution of isolated
massive binaries (where the components have initial masses $\geq 8 M_\odot$) or
can be formed dynamically (i.e., in dense globular and nuclear star clusters) from binary-single
star interactions between compact remnants and primordial binaries \citep{Mandel:2009nx}. 
While the key stages of the binary evolution are well understood,
there are significant uncertainties in the details of the process. 
Population synthesis codes attempt to model these uncertainties using empirical prescriptions. 
These models contain numerous parameters which are not well constrained relating to astrophysics 
such as stellar winds, supernova kicks imparted on black holes at birth and common envelope binding 
energy among others. Varying these parameters will have a significant impact on both the predicted 
rate of compact binary mergers, as well as the distribution of expected masses and spins of the compact remnants that comprise the binary \citep{Dominik:2012kk}.  

In this paper, we introduce a straightforward model selection method to
distinguish between various formation and evolution scenarios.  We focus on the
two parameters that will be best measured: the overall rate of binary mergers
and the chirp masses of the observed binaries.  

Furthermore, we restrict attention to \acp{BBH} as, based upon the recent population synthesis models,
these are predicted to be the most numerous
\citep{2003MNRAS.342.1169V,Dominik:2012kk}. We caution, however, that detection
rates are highly uncertain and previous papers have argued that there will be
essentially no \acp{BBH} \citep{0004-637X-662-1-504,2014A&A...564A.134M}.
\change{This trivially means that any detections of merging \acp{BBH} will allow
models predicting a dearth of such systems to be ruled out, shedding light on
the astrophysical assumptions made therein. Beyond that,} we
show how, in addition to the merger rates, the broad range of \BBH chirp masses
predicted by population synthesis models encodes information about the \BBH
formation mechanisms.                                          

\change{There have been many studies performed over the last decade that have
made use of either one or both of these pieces of information to distinguish
between competing astrophysical models. \cite{1538-4357-589-1-L37} used a
Kolmogorov-Smirnov test to compare simulated \GW chirp mass measurements to
a series of predicted observed distributions from population synthesis models.
They find they can distinguish many models with $\sim 100$ observations, a
finding we confirm in the present study. 
\cite{2041-8205-725-1-L91} use a Bayesian approach introduced in
\cite{PhysRevD.81.084029} to show how one can use \GW observations along with
dark matter simulations to distinguish between different natal kick-velocity
models, and again find they require $\mathcal{O}(100)$ observations to
distinguish between models.

\cite{2012arXiv1208.0358B} discuss using  upper limits on binary merger rates to
distinguish between population synthesis models.
Recently, \cite{2015MNRAS.450L..85M} have shown how one can use population
synthesis models along with \GW observations of binary mergers to measure the
relative rate of \BNS, \NSBH and \BBH mergers with $\mathcal{O}(10)$
observations. In addition,
\cite{2013NJPh...15e3027M} show how one should use all of the information available to avoid selection biases when attempting to make inferences about distributions of rates and parameters of merging binaries.

More sophisticated techniques have also been discussed in the literature.
\cite{PhysRevD.88.084061} introduces a framework to incorporate measurements of
both the merger rate and parameter distributions of \GW observations, and
compares these to a set of population models which sparsely sample the relevant
parameter space. A similar technique is used in \cite{Mandel:2009nx} (see also
\cite{2010arXiv1001.2583M}).

Here, we introduce a fast, simple method to make inferences about
astrophysical models using information from \GW observations.  
The method is general, and could be applied to any set of binary evolution
models.  We illustrate its utility by evaluating the ability to
distinguish between a suite of population synthesis models \citep{Dominik:2012kk}.
For concreteness, we restrict attention to the expected results from the early observing
runs of the advanced \GW detector era \citep{Aasi:2013wya}.
}

Population synthesis models typically predict the galactic rate of binary
mergers and the parameter
distributions.  From this, we model the observed distribution by accounting for observational bias: 
\GW detectors are able to observe signals from higher mass systems to a greater
distance.
Additionally, we incorporate cosmological effects that lead to a red-shifting of both the observed
masses and the observed merger rate.  Finally, we model measurement errors and uncertainties
inherent in the extraction of the signal from a noisy data stream.  For each population synthesis
model, we generate an expected observed rate and associated mass distribution.  

Based on simulated observational results, we can use model selection to differentiate between the various
models.  To give a sense of what we can expect, we simulate results from the early \aLIGO
and \AdV observational runs.  To do this, we choose one of models from a suite of population synthesis models to play
the role of the universe, and draw \GW observations of \acp{BBH} from it, 
accounting for known observational biases and anticipated measurement errors. 
We then compare these observations to the full suite of population synthesis models
and, starting with a uniform prior on the models, we compute the posterior
probability for each model.  

While the results that we present are limited to these specific scenarios, the method we
introduce is general and could easily be applied to the predictions from any population
synthesis model and the results (predicted or observed) from any detector network. 
We also caution the reader that the models of \dominik represent the most optimistic 
predictions of \BBH merger rates, with all models predicting a detection within the first 
two \aLIGO and \AdV science runs. Lower merger rates would lead to observations of \BBH 
mergers only in later runs at, or close to the design sensitivity of the detectors. For an overview of rate 
predictions for \aLIGO and \AdV see \cite{Abadie:2010cf}.

This paper is structured in the following way. In
Section~\ref{sec:binary_evolution} we give a brief review of compact binary
formation, and introduce the models we use in Section~\ref{subsec:models}. In
Section~\ref{sec:observationalbias} we describe our algorithm for accounting for
known selection biases, converting an intrinsic chirp mass distribution to a
predicted observed distribution. Section~\ref{sec:calculating_probability} shows
how to use information from the two well measured parameters --- the chirp mass
and the merger rate --- to distinguish between population synthesis models. In
Sections~\ref{sec:observing_scenarios} \& \ref{sec:results} we show what we may
be able to learn about binary evolution using \GW observations of binary black
holes from the first two \aLIGO and \AdV science runs. Finally in
Section~\ref{sec:conclusion} we conclude and suggest areas which require further
investigation.

\section{Compact binary formation and evolution}
\label{sec:binary_evolution}

In this section, we provide a brief review of isolated binary evolution,
highlighting the poorly understood stages of the evolution, which 
lead to the uncertainties in the predicted merger rates and mass distributions of
the binaries. For more information see a review such as \citet{lrr-2006-6}. 

\subsection{General overview}

For a single star, its evolution is solely determined by the zero-age main
sequence (ZAMS) mass and composition. However, the majority of massive stars
exist in binaries or multiple systems, with $\gtrsim 70 \%$ of massive O-type
stars exchanging mass with a companion during their lifetime
\citep{2013A&A...550A.107S,1991A&A...248..485D}. In this
case, the evolution is no longer straightforward, and can lead to a plethora of
exotic systems. Here we give one possible evolutionary pathway for a massive
binary; many alternative pathways also exist (see for example Tables~4 \& 5 in
\citet{Dominik:2012kk} for a summary).

Consider a binary in which both stars have ZAMS masses $\gtrsim 8 M_\odot$. 
The initially more massive star (the primary) in the binary
will evolve off of the main sequence first since it has the shorter lifetime. As
it evolves, its radius expands until it fills its Roche Lobe as a giant and
begins to transfer mass to the companion (the secondary) star, stripping the primary's hydrogen outer layers and leaving a 
He/Wolf-Rayet star. Already the evolution of the binary is different to that of
single stars since the companion can change its mass considerably, leading in
some cases to a reversal of the mass ratio. If the core is massive enough, the primary will then collapse in a
supernova, and leave behind a compact remnant --- either a neutron star or a
black hole depending on the pre-supernova core mass. 

In stellar evolution models, the distinction between collapse to a neutron star
or a black hole is made via mass alone, with the maximum allowed mass of a
neutron star being one of the free parameters. In reality, the maximum neutron
star mass is set by the unknown neutron star equation-of-state. The maximum
observed neutron stars have masses around $2 \Msun$ \citep{Demorest:2010,2013Sci...340..448A}. Causality
and General relativity require the maximum neutron star mass to be $\leq 3.2
\Msun$ \citep{PhysRevLett.32.324}.

The mechanism of the supernova itself is intensely studied but still not fully
understood. If the supernova is asymmetric (due to asymmetric mass loss or
neutrino emission) the resulting neutron star can be given a natal kick velocity
due to the conservation of momentum, which is of the order $250 \text{ km
s}^{-1}$ for galactic neutron stars \citep{2005MNRAS.360..974H}. It is unclear
whether black holes also receive a kick of this magnitude or whether mass
falling back onto the black hole reduces the size of this kick significantly (see for e.g. \cite{2012MNRAS.425.2799R,2013MNRAS.434.1355J}).

If the system survives the first kick, then the secondary will begin to evolve.
The compact remnant 
accretes matter from the stellar wind of its companion, becoming a luminous X-ray source. At this 
stage, the binary may be observable electromagnetically as a high-mass X-ray binary. Although 
the theory of stellar winds is fairly  robust \citep{CAKstellarwinds}, the strength of stellar winds in these systems remains uncertain \citep{2008AJ....136..548L}.

As the secondary continues to evolve, it will continue to expand and fill its Roche Lobe. If the mass transfer through Roche Lobe Overflow is unstable, a common envelope phase \citep{2013A&ARv..21...59I,1976IAUS...73...75P} can be initiated. This is where both the compact remnant and the core of the secondary orbit within the secondary's hydrogen outer layers. The common envelope is the least well understood phase in the evolution of binaries. The common envelope is usually parametrised in one of two fashions; the $\alpha$ prescription \citep{1984ApJ...277..355W} focusing on conservation of energy, or the $\gamma$ prescription \citep{Nelemans:2000xs} focusing on conservation of angular momentum. The core and compact object spiral in towards one another on a dynamical timescale due to drag, and orbital energy is used to eject the envelope. This stage is responsible for dramatically reducing the orbital separation in the binary.

If the binary survives the common envelope, the core of the secondary can then
go supernova, potentially imparting a second kick on the system (although it is
generally less likely to unbind the system since the orbital velocities are now
much higher). Finally, a compact binary remains containing neutrons stars and/or
black holes. It is these systems which then inspiral towards one another and
merge due to radiation reaction, and will be observed in \acp{GW} by \aLIGO and
\AdV.

\subsection{Detailed binary evolution models}
\label{subsec:models}

Population synthesis codes are Monte-Carlo simulations that evolve large ensembles of primordial binaries via semi-analytical prescriptions, taking as input parameters corresponding to the poorly understood astrophysical stages outlined above. Binary population synthesis models can be used to try to understand the effects of these uncertainties on binary evolution, and on the resultant population of compact binaries. One way to exploit the information contained in \GW observations of coalescing \acp{BBH} is therefore to compare the measured properties of a population to population synthesis model predictions.

For this study we use a set of publicly available\footnote{\url{http://www.syntheticuniverse.org}} population synthesis models presented in \citet{Dominik:2012kk}, produced using the \texttt{StarTrack} population synthesis code \citep{Belczynski:2005mr}. Predicted chirp mass distributions and merger rates of \BNS, \NSBH and \BBH systems are provided for a range of input physics.  

The relative rates of \BNS, \NSBH and \BBH mergers are uncertain.  Although \BBH systems are more massive
(and consequently detectable to a greater distance), much more massive stars are needed in order to form them, 
and the \IMF falls off rapidly at high masses, meaning these stars are rarer. It is also worth noting that no \BBH has ever
been observed, although systems which may be progenitors for them such as Cyg
X-3 \citep{Belczynski:2012jc}, IC 10 X-1 \citep{Bulik:2008ab} and NGC 300 X-1 \citep{Crowther01032010} have been studied and provide some limits on \BBH merger rates.
The population synthesis model we are utilising predicts that \BBH detection rates will dominate over 
those for \BNS and \NSBH.  Based on this, and to keep
the analysis simple, we restrict our attention to \BBH systems. It would be relatively straightforward to extend
the framework we introduce to include \textit{all} \GW observations of binary
mergers.

\begin{deluxetable}{l p{6cm}}
 \tablecolumns{2}
 \tablecaption{Summary of population synthesis models.}
\tablehead{ \colhead{Model} & \colhead{Physical difference}}
\startdata
Standard  	& Maximum neutron star mass = 2.5 \Msun,
\emph{rapid} supernova engine \citep{Fryer:2011cx}, physically motivated
envelope binding energy \citep{2010ApJ...716..114X}, standard kicks
$\sigma = 265 ~\text{ km s}^{-1}$					\\
		Variation 1 	& Very high, fixed envelope binding energy\tablenotemark{a} 
\\          
		Variation 2 	& High, fixed envelope binding energy\tablenotemark{a} 	
\\
		Variation 3 	& Low, fixed envelope binding energy\tablenotemark{a}
\\
		Variation 4 	& Very low, fixed envelope binding energy\tablenotemark{a} 
\\
		Variation 5 	& Maximum neutron star mass = 3.0 \Msun 
	\\
		Variation 6 	& Maximum neutron mass = 2.0 \Msun 		
\\
		Variation 7 	& Reduced kicks $\sigma = 123.5 ~\text{ km
s}^{-1}$ \\
		Variation 8 	& High black hole kicks, $f_b = 0$		
\\
		Variation 9 	& No black hole kicks, $f_b = 1$ 		
\\
		Variation 10 	& \emph{Delayed} supernova engine
\citep{Fryer:2011cx} 	\\
		Variation 11 	& Reduced stellar winds by factor of 2 
\enddata
\tablenotetext{a}{See Section~2.3 in \dominik for details}
\tablecomments{Models presented in \citet{Dominik:2012kk}, with parameter
variations indicated in the second column which broadly relate to the
uncertainties in binary evolution discussed in the text. All other parameters
retain their standard model value.\label{tab:models}}
\end{deluxetable}

We use the set of 12 population synthesis models for which predicted rates and mass distributions are available.
These models are summarised in Table \ref{tab:models}.  The standard model assumes a maximum neutron star 
mass of $2.5 \Msun$, uses the \emph{rapid} supernova engine detailed in \citet{Fryer:2011cx}, physically motivated
common envelope binding energy \citep{2010ApJ...716..114X}, and kick velocities for supernova remnants drawn 
from a Maxwell distribution with a characteristic velocity of $\sigma = 265 ~\text{ km s}^{-1}$. 
There are then eleven variations, in each of which one of the above parameters is varied: the first four
variations consider changes in the energetics of the common envelope phase, the next two vary the maximum mass
of neutron stars, three more change the kick imparted on the components during collapse to a neutron star or black hole
and the final two consider a delayed supernova engine and a change in the strength of stellar winds.  The models
are described in detail in section 2 of \dominik.

We expect that in general, the true universe will not match one of a small set
of models, but will lie in between these models (or potentially outside of them
if additional unmodelled physics is required to accurately describe binary
evolution). Assumptions that are not varied in these models, but which may have
a large impact on the resultant \BBH distribution include distributions of the
parameters of primordial binaries (\IMF, orbital elements \citep{Belczynski:2015in_prep}), tides, stellar rotation \citep{2013ApJ...764..166D} and magnetic fields. Here we neglect these additional considerations and investigate how one can differentiate between a small suite of population synthesis models using \GW observations of \acp{BBH}. 
\change{A full treatment of these additional properties has the potential to significantly impact
 stellar evolution models and may well lead to degeneracies whereby significantly
 different astrophysical models predict comparable populations of binaries.}

\change{
Since calculating population synthesis models can be computationally expensive, the models 
are discretely sampled over a large range of parameter space (in some
cases orders of magnitude) in an attempt to bracket the truth. Furthermore, each of the
models used in this study varies only one parameter from its standard value at a
time. It is quite likely that the true values of many of these parameters will differ from
those presented in \dominik, resulting in a population that does not match any of the 
ones included here. Varying combinations of parameters will also need to be studied, 
as this may lead to issues with degeneracies in which combinations of parameters can be 
determined from \GW observations. To be able to reliably extract the details of stellar
evolution from \GW observations, one would require to have models
calculated on a dense enough grid that one can perform interpolation between
them \citep{PhysRevD.88.084061,O'Shaughnessy:2006wh}. 
}

\subsubsection{Metallicity}
\label{subsubsec:met}

Each model is calculated at solar ($Z = 0.02 = Z_\odot$) and sub-solar ($Z =
0.002 = 0.1 Z_\odot$) metallicities. In addition, there are two 
submodels that differ in the way binaries entering into a common
envelope when one of the stars is on the Hertzsprung gap are handled (see
section~\ref{subsubsec:hg}).

We choose to use a 50-50 mixture of the solar and sub-solar models as used in
\citet{Belczynski:2010tb}, motivated by results from the Sloan Digital Sky
Survey (SDSS, \citep{Panter11122008}) showing that star formation is
approximately bimodal with half of the stars forming with $Z \sim Z_\odot$ and
the other half forming with $Z \sim 0.1 Z_\odot$. For the future, it would be
desirable to include a more thorough treatment of the metallicity distribution,
including its evolution with cosmic star formation history as done in
\citet{2013ApJ...779...72D,2014arXiv1405.7016D}. 

Although metallicity in the local universe may be bimodal, one still expects a smooth distribution of metallicities to exist. Using only a discrete mixture of solar and sub-solar metallicity predictions may give rise to non-physical peaks or sharp features in the chirp mass distributions which may artificially aid in distinguishing between them \citep{2014arXiv1405.7016D}. However, in practise we find that these peaks are sufficiently smoothed out by measurement errors (see section~\ref{subsec:errors}).

\change{

Studies have shown that the majority of \acp{BBH}
observable by \aLIGO were formed within $\sim 1 \text{ Gyr}$ of the Big Bang
\citep{2013ApJ...779...72D,2014arXiv1405.7016D}, when the metallicity of the
universe was distinctly lower. This is due to a number of reasons (see
for example \cite{Belczynski:2010tb}). It is easier for supernova
progenitor stars to remain massive at lower metallicities due to weaker stellar
winds compared to at solar metallicity. Also, many potential \BBH progenitor
systems merge prematurely at higher metallicities during the CE phase since the
secondary is likely to be on the Hertzsprung Gap, whereas at lower metallicities
the secondary does not expand enough to initiate a CE event until it is a
core-helium burning star (see \citet{Hurley01072000} for the effect of
metallicity on stellar radius). These \acp{BBH} are formed with long delay times
such that they are only merging now. One therefore needs to include the time
evolution of metallicity to accurately model the expected population of \acp{BBH}
mergers \citep{2013ApJ...779...72D}.
}

\subsubsection{Fate of Hertzsprung Gap donors}
\label{subsubsec:hg}

The Hertzsprung gap is a short lived (Kelvin-Helmholtz timescale)
stage of stellar evolution where a star evolves at approximately constant
luminosity across the Hertzsprung-Russell diagram after core hydrogen burning
has been depleted but before hydrogen shell burning commences.

Whilst on the main sequence, stars are core burning hydrogen, and do not possess
a core-envelope separation as the helium core is still being formed. Therefore, if a main sequence star enters into a common envelope, orbital energy is dissipated into the whole star, rather than just the envelope,
making ejecting the envelope extremely difficult. It is therefore expected that
any star entering into a common envelope phase whilst on the main sequence will
result in the two stars merging prematurely in an event which is not interesting
from a \GW astronomy point of view. 

For stars that are on the Hertzsprung gap, the situation is not so clear. The helium core begins contracting whilst the envelope of the star expands. It is unclear if there is sufficient core-envelope separation on the Hertzsprung gap for a star entering a common envelope phase to have its envelope ejected, or whether it would suffer a similar fate to a main sequence star.

The fate of Hertzsprung Gap donors is another of the uncertainties that is investigated by \dominik. 
In the optimistic submodel (referred to as submodel A in
\citet{Dominik:2012kk}), the authors ignore the issue and calculate the common
envelope energetics as normal \citep{1984ApJ...277..355W}. In the pessimistic
submodel (referred to as submodel B), any binary in which the donor is on the
Hertzspung gap when the binary enters into a common envelope phase is assumed to
merge. This tends to reduce the number of merging binaries (and thus the rates)
compared to the optimistic model. It is unlikely that either of these models is accurate, 
as the fate of a Hertzsprung gap donor will depend on the internal structure of the star 
as it enters the common envelope phase.  Nonetheless, submodels A and B provide upper
and lower bounds, respectively, on the number of Hertzsprung gap donors forming \BBH.

In this paper, we compare results for the twelve models listed above for \textit{both}
the optimistic (submodel A) and pessimistic (submodel B) Hertzsprung gap evolution.

\section{Predicted observed distributions}
\label{sec:observationalbias}

For each of the models described above, we are given an expected rate of
binary mergers per \MWEG, as well as a distribution of binary parameters
(notably the component masses).
The population of \acp{BBH} observed by the advanced \GW detectors
will differ from this underlying intrinsic distribution due to the following observational effects.
\begin{enumerate}[(a)]
 \item The \GW signal from binaries at large distances will be red-shifted due
to the expansion of the universe which consequently leads to a shifted
measurement of the binary's total mass. 
 \item The \GW amplitude scales with the binary's total mass, thus binaries
with heavier components will be observable to greater distances, provided their signal still lies
in the sensitive frequency region of the detector, which leads to an increased
number of observed high-mass systems.
 \item Due to the presence of noise in the detector the best-measured
parameters will differ from the binary's intrinsic parameters which effectively
blurrs the observed distribution.
\end{enumerate}%
We take all three effects into account and calculate the distribution of
parameter we expect to observe. \change{Our approach is
consistent with previous methods in the literature (e.g.,
\citet{Dominik:2014yma}), apart from how we account for
measurement errors across the parameter space. For
completeness, in the remainder of the section, we briefly 
recap how these effects are accounted for and the observed distribution 
obtained.}

\subsection{Detectability}
\label{subsec:chirp mass}

We model the \GW signals by the dominant harmonic only, which is sufficient for
the majority of black hole systems we are considering \citep{Capano:2013raa,
Bustillo:2015ova}. The signal observed in a
\GW detector can then be expressed as
\citep{Fairhurst:2007qj}
\begin{equation}
  h(t) = \frac{1}{D_{\mathrm{eff}}} \left[ h_{0}(t) \cos\Phi +
h_{\pi/2}(t) \sin \Phi \right],
\end{equation}
where $D_{\mathrm{eff}}$ is called the effective distance, $\Phi$ is the
coalescence phase as observed in the detector and $h_{0,\pi/2}$ are the two
phases of the waveform which are offset by $\pi/2$ relative to each other
[equivalently, their Fourier transforms obey $\tilde h_{0}(f) = i
\tilde h_{\pi/2}(f)$].
The effective distance is defined as
\begin{equation}
D_{\mathrm{eff}} = \frac{D_L}{\sqrt{ F_{+}^{2} (1 + \cos^{2} \iota)^2/4 +
F_{\times}^{2} \cos^{2} \iota}}.
\end{equation}
$D_L$ is the luminosity distance to the binary, $F_{+, \times}$ are the detector
response functions and $\iota$ is the binary inclination angle.  The maximal
(and circularly
polarised) \GW signal is obtained when the signal is directly overhead the
detector $F_{+} = 1$; $F_{\times} = 0$ and  with $\iota = 0, \pi$ corresponding
to a face on signal.

The effective distance is inversely proportional to the \SNR, $\rho$, which is
defined as \citep{Cutler:1994ys,Poisson:1995ef}:
\begin{equation}
\rho^2 = 4 \int_{f_{\rm low}}^{\infty} \frac{\vert \tilde{h}(f) \vert^{2}}{S_n
(f)} df,
\label{eq:snr_def}
\end{equation}
where $\tilde{h}(f)$ is the frequency-domain gravitational waveform and the
detector noise power spectral density is denoted by $S_n (f)$. \change{We choose
a lower cutoff frequency of $f_{\rm low} = 20\,\textrm{Hz}$, suitable for the early advanced detectors.} 
The \SNR at which a signal can be detected will depend upon the details of the
detector network, including the sensitivities of the detectors as well as the
character of the data --- non-stationarities in the data make it more difficult
to distinguish candidate signals from the background noise. However, for studies such as this,
it is convenient to choose an approximate threshold.  Experience has shown that
a network \SNR of 12 is approximately where we might expect to make a detection
\citep{Colaboration:2011np, Aasi:2013wya}.  This corresponds to an \SNR of
around 8 in each of the LIGO detectors in the early science runs\footnote{For
the early science runs, we expect the LIGO detectors to be about twice as
sensitive as Virgo so, on average, one might expect a threshold event to have
\SNR of 8 in each of the LIGO detectors and 4 in Virgo}. For the studies
presented in this paper, we use this simple, single detector threshold to decide
whether a signal would be observed by the detector network.  

Given a model for the waveform, $h(t)$, we can calculate the maximum
effective distance to which the signal could be detected.  This is known as the
\textit{horizon distance}, $D_H$, and corresponds to the maximal distance at
which the signal could be observed if it is optimally oriented and overhead.
To calculate the horizon distance we use the
phenomenological waveform model introduced by \citet{Santamaria:2010yb} that
includes the inspiral, merger and ringdown sections of the
waveform calibrated using numerical relativity.  The model provides the waveform
$\tilde h(f)$ in the frequency domain as a function of the binary's total mass
$M$, its symmetric mass ratio $\eta$ and an effective total spin parameter,
$\chi$.

The mass parameters of the binaries are characterized in terms of 
the best measured parameter combination, the so called
\emph{chirp mass} $\mathcal M$, which is a combination of the
component masses $m_1$ and $m_2$,
\begin{equation}
\mathcal{M} = \frac{(m_1 m_2)^{3/5}}{(m_1 + m_2)^{1/5}} = M \eta^{3/5},
\label{eq:Mchirp_def}
\end{equation}
where $M = m_1 + m_2$ is the total mass, and $\eta$ is the symmetric mass ratio,
\begin{equation}
\eta = \frac{m_1 m_2}{(m_1 + m_2)^2} \leq 0.25.
\end{equation}
For an equal mass binary $m_1 = m_2 = m$, the chirp mass $\mathcal{M} \approx
0.87 m$. \change{In the remainder of the paper, we will focus on the predicted and
observed chirp mass distributions, and not consider mass ratio or spin.}

Our aim is to predict the observed chirp mass distribution given the
intrinsic model prediction, and compare these with observations.

\begin{figure}
	\centering
	\includegraphics[width=\linewidth]{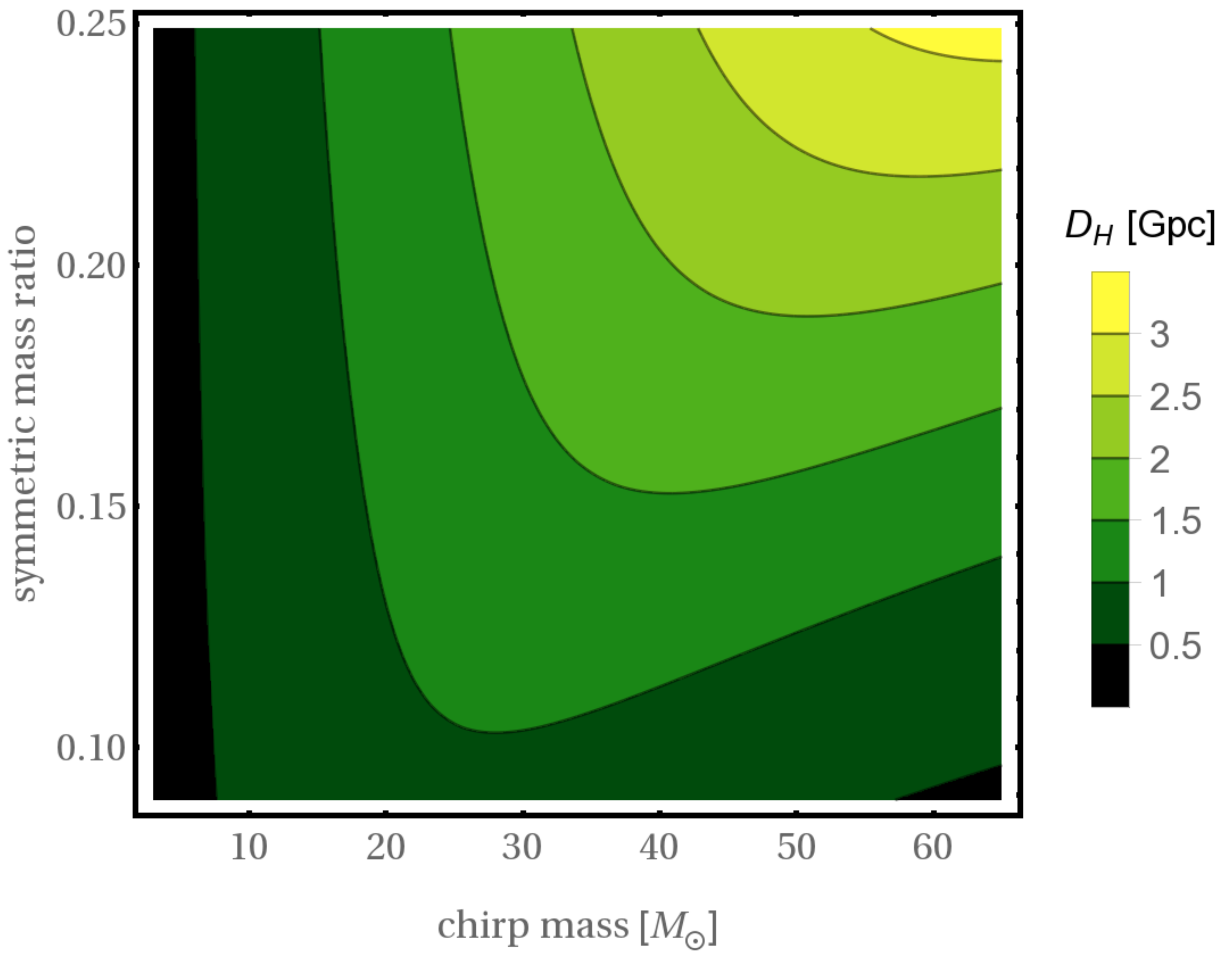}

	\caption{Horizon distance in Gpc for nonspinning \acp{BBH} as a
function of chirp mass and symmetric mass
ratio assuming a single detector with the early \aLIGO noise spectrum.}
	\label{fig:horDistPlot}
\end{figure}

Throughout most of this paper, we assume the early \aLIGO (circa 2015) noise
spectrum \citep{earlyaLIGO,Aasi:2013wya} representing the expected 
sensitivity of \aLIGO during its first observing runs.
A plot of the horizon distance as a function of the chirp mass and the
symmetric mass ratio is given in Figure~\ref{fig:horDistPlot}. It encodes the
farthest distance to which a \BBH with the given parameters can be seen.
\change{The horizon distance will also be a function of the black hole spins.
Since \citet{Dominik:2012kk} do not provide individual spin information in their
catalogues, we set the spin parameter to zero for simplicity when simulating
signals in our synthetic universe.  Our ignorance of the spins may lead to systematic biases, 
as high spins can noticeably affect the horizon distance \citep{Ajith:2009bn} and could
change the rate of observed signals by a factor of two or three \citep{Dominik:2014yma}. One
could incorporate this lack of knowledge by assuming a spin distribution for black holes
and margnializing the result over the spins.  We defer this to a
later study when more informed spin priors (observationally motivated or from
population synthesis) can be incorporated.}

Not every binary within the horizon will be detected, as $D_{\rm eff}$ is
location and orientation dependent. Under the assumption of a
uniform distribution over the sky and a uniform source orientation, however, we
can numerically calculate the fraction $P(\xi)$ of systems with
\begin{equation}
 \frac{D_L}{D_{\rm eff}} = \sqrt{ F_{+}^{2} (1 + \cos^{2} \iota)^2/4 +
F_{\times}^{2} \cos^{2} \iota} > \xi,  \label{eq:P_def}
\end{equation}
with $\xi \in [0,1]$. Note that $P(\xi)$, which we can interpret as a
cumulative distribution
function, is independent of the binary's masses, and we will use it
to determine what fraction of signals at a given luminosity distance is
detectable, i.e., has an \SNR larger than the detection threshold. 

\subsection{Cosmological effects}

We simulate an expanding universe with sources distributed uniformly and
isotropically in comoving volume, which on scales of hundreds of Mpc is a
valid assumption. Since the frequencies of any signal become increasingly
redshifted with growing distance between source and detector, the total chirp mass 
measured at the detector is shifted by
\begin{equation}
 \mathcal M^\ast = \mathcal M \, ( 1+z), \label{eq:Mc_redshift}
\end{equation}
where $z$ denotes the redshift. Assuming zero curvature and standard
cosmological parameters \citep{Bennett:2014tka}
\begin{equation}
 \Omega_M = 0.286, \quad \Omega_\Lambda = 1 - \Omega_M, \quad
 H_0 = 69.6,
\end{equation}
we calculate the comoving distance as a function of the redshift \citep{Hogg:1999},
\begin{equation}
 D_C(z) = \frac{c}{H_0} \int_0^z \frac{dz'}{\sqrt{\Omega_M (1+z')^3 +
\Omega_\Lambda}}.
\end{equation}
Here, $c$ denotes the speed of light. 

The catalogues by \citet{Dominik:2012kk} provide large sets of binaries
characterized by their intrinsic chirp mass $\mathcal M$ and symmetric mass
ratio $\eta$. When we distribute them uniformly in \change{comoving volume}, the
observed
chirp masses $\mathcal M^\ast$ are redshifted according to
\eqref{eq:Mc_redshift}. This implies that the maximal distance to which they can
be detected changes as it is the observed parameters, not the
intrinsic parameters, that determines the appropriate horizon distance. Since
\begin{equation}
 D_L = D_C (1 + z), 
\end{equation}
the maximal observable comoving distance satisfies
\begin{equation}
 D_C^{\rm max}(\mathcal M, \eta, z) \; (1 + z) = D_H(\mathcal M^\ast, \eta), 
\label{eq:DCmax}
\end{equation}
which we solve numerically for $z$. Note that the leading-order inspiral
horizon distance behaves as $D_H(\mathcal M^\ast) \sim (\mathcal
M^\ast)^{5/6}$, hence
\begin{equation}
 D_C^{\rm max}(\mathcal M) \sim \frac{\mathcal M^{5/6}}{(1+z)^{1/6}} <
D_H(\mathcal M). \label{eq:max_DC}
\end{equation}
While the derivation of \eqref{eq:max_DC} is only valid for low-mass systems, we
find that $D_C^{\rm max}$ is generally less than the static, Euclidean universe equivalent
$D_H$.

\subsection{Detection rate and distance distribution}
\label{subsec:rates}

We now assume binaries of a fixed model, distributed isotropically and uniformly
in comoving volume, that merge at a constant comoving merger rate density
$\mathcal R$ (in $\text{MWEG}^{-1}\;\text{Myr}^{-1}$) as given in the data sets
by \citet{Dominik:2012kk}. To convert these numbers into a detection rate for
\aLIGO, we proceed as follows: 

First, the comoving merger rate $\mathcal R$ per \MWEG has to be multiplied 
by an average galaxy density which we take
as $\rho_G \approx 0.0116\, \text{Mpc}^{-3}$ following
\citet{0004-637X-675-2-1459}. 
Next, we must calculate the effective volume in which
each binary is observable, by integrating the number of observable
binaries as a function of $D_{C}$.  As the distance increases, 
the area of the corresponding sphere increases as $D_{C}^{2}$ but 
the fraction of binaries that are oriented such that their signal is
sufficiently loud for detection (that is, $D_{\rm eff} < D_H$) becomes
smaller. 
Finally, due to the redshifted time, 
\begin{equation}
 t_L = t_C (1+z)
\end{equation}
the observed merger rate for binaries at redshift $z > 0$ is less than the
comoving merger rate. Thus, the effective volume for a binary with chirp mass
$\mathcal{M}$ is
\begin{equation}
 V_{\rm eff}(\mathcal{M}) = 4\pi \int_0^{D_C^{\rm max}} \frac{D_C^2}{1+z} \;
P\!\left(\frac{D_L}{D_H(\mathcal
M^\ast, \eta)} \right) dD_C \, , \label{eq:Veff}
\end{equation}
where
$D_C^{\rm max}$ is defined by \eqref{eq:DCmax}. The function $P$, introduced in 
Eq.~\eqref{eq:P_def}, gives the fraction of suitably
oriented binaries (i.e., those giving an SNR greater than 8)  and
$(1+z)^{-1}$ accounts for the difference between apparent and comoving merger
rate density. We note that the integrand in \eqref{eq:Veff} can be
interpreted (up to a normalisation) as the observed distance distribution for
binaries with fixed intrinsic parameters.

The average detection rate for each model is given by
\begin{equation}
 \dot N = \mathcal R \times \rho_G \times \overline{V_{\rm eff}},
\label{eq:det_rate}
\end{equation}
where $\overline{V_{\rm eff}}$ denotes the average effective volume, with
the average taken over all binaries in a given model.
We take $\mathcal R$ and $\rho_G$ from \citet{Dominik:2012kk} and
\citet{0004-637X-675-2-1459}, respectively. 

\begin{figure}
 \centering
 \includegraphics[width=\linewidth]{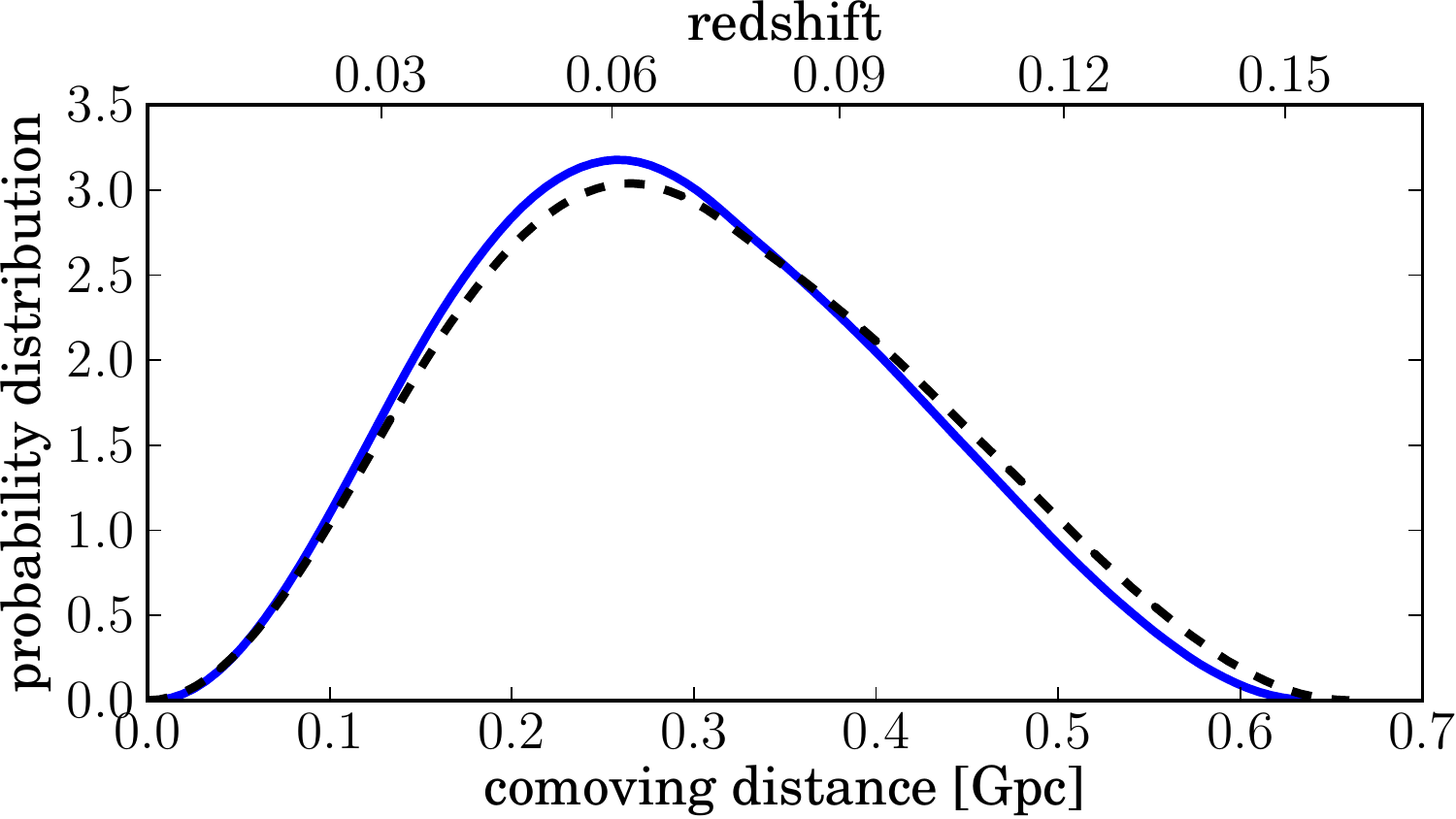}
 \caption{The probability distribution in comoving distance for detectable
\acp{BBH} with $m_1 = m_2 = 10M_\odot$. The solid (blue online) curve takes
cosmological effects into account (see text) while the dashed line assumes a
static, Euclidean universe (i.e. local universe approximation).}
\label{fig:dc_dist}
\end{figure}

Figure~\ref{fig:dc_dist} shows this distribution for an
equal-mass \BBH with $m_1 = m_2 = 10M_\odot$. For comparison, we include the
equivalent curve for a static, Euclidean universe, where $D_L \equiv D_C$ and
\eqref{eq:Veff} simplifies to the case $z=0$. As expected, both curves
agree for low redshift, but as we have noted above, there are fewer binaries
seen at large comoving distances if the expansion of the universe is taken into
account. This effect becomes increasingly important for larger distances, i.e.,
for high-mass binaries and more sensitive detector configurations.

The effective volume in which binaries with fixed parameters are detectable
changes considerably across the parameter space. This leads to an observational
bias in favour of systems with large volume reach. 
We incorporate this effect by re-weighting the chirp mass distribution of
binaries according to their individual effective volumes.
In practice, \citet{Dominik:2012kk} provide the data for each of
their models in form of a discrete set of binary parameters. For each of those
binaries, we calculate the integer part of $V_{\rm eff}/ V_{\rm eff}^{\rm min}$
and add that many copies of the binary to our new set of observable parameters.
Here, $V_{\rm eff}^{\rm min}$ denotes the smallest effective volume across all
binaries in the set, and only one copy of the binary with this smallest
effective volume is kept.\footnote{We could keep more copies of the binary with
the smallest effective volume and multiply the number of every other binary in
the set accordingly, but tests showed that this has no effect on our results.}

Finally, for each binary in our new set, we draw a comoving distance from the
distribution underlying $V_{\rm eff}$. From this distance, we then infer the
redshift and change from $\mathcal M$ to the observable redshifted chirp mass
$\mathcal M^\ast$ according to \eqref{eq:Mc_redshift}. Our discrete
representation of observable binaries then consists of multiple copies of the
same intrinsic systems, each however with a unique redshifted chirp mass. 

Note
that an equivalent, but computationally more expensive, procedure would be to
randomly draw binaries from the intrinsic distribution, then draw 
comoving distances and orientations for each binary within the total sensitive
volume for the respective model and only keep those binaries with a detectable
\GW signal. Our method instead avoids disregarding any randomly drawn sources by
drawing from the appropriate (distance/orientation) distribution of detectable
signals.

\subsection{Estimating and Including Measurement Errors}
\label{subsec:errors}

Including the observational bias discussed in Sec.~\ref{subsec:rates} in the
chirp mass distribution still does not yield the distribution that one would
expect to observe, because there will be a measurement error associated with
each of the observations. \change{Previous publications have mainly discussed a
full Bayesian framework to combine multiple observations including their
measurement uncertainties
\citep{PhysRevD.81.084029,Mandel:2009nx,PhysRevD.88.084061}. We instead assume a
statistical fluctuation of the measured parameter around its true value as
detailed below.}

The accuracy of the parameters recovered during \GW searches is
limited by two factors. First, since we match to templates
of the signals, the accuracy of recovered parameters will be limited by the
accuracy of the waveform models that used in the search. Second, the accuracy
will
be affected by statistical fluctuations of the noise in the measurement process.
While we leave the former for dedicated studies such as
\citet{Buonanno:2009zt} and \citet{Nitz:2013mxa}, we can estimate the
uncertainty due to the latter using the well-known Fisher matrix estimate.

Fisher matrix analyses rely on a linear approximation of signal variations and
are valid for large \acp{SNR}. Neither of the two assumptions is typically
valid in realistic scenarios, and recent papers have discussed some
implications of violating these assumptions \citep{Vallisneri:2007ev,
Rodriguez:2013mla, Mandel:2014tca}. Here, however, in order to demonstrate the
basic efficacy of our method to distinguish \BBH populations with \GW
observations, we take Fisher-matrix predictions as a proxy for 
\change{the width of posterior
distributions of parameters obtained via a fully Bayesian analysis of the kind that will be performed on actual \GW events
\citep{Veitch:2014wba}. In performing a population study of the kind we perform here, one should include not only a point estimate for parameters such as the chirp mass, but the full posterior from these parameter estimation routines. These posteriors can then be combined in the correct way, 
as described in \cite{PhysRevD.81.084029}. The method we use here is essentially 
the point estimate approximation to the full analysis.}

\change{We employ the same inspiral-merger-ringdown model
\citep{Santamaria:2010yb} for our Fisher-matrix calculations as we used
to simulate \GW signals. We only consider variations of the intrinsic
parameters: masses, time, phase and a model-specific single effective spin.}
We assume that these are also the parameters that are recovered, at
least
initially by the \GW search algorithm (see, e.g., the
recently proposed search algorithm for nonprecessing, spinninng
binaries by \citet{Canton:2014ena}). This assumption is
likely to make our error estimates too large since actual \GW events will be
followed up by complex parameter estimation routines (see e.g.,
\citet{Veitch:2014wba}) exploring the full parameter space of precessing
binaries \citep{Vitale:2014mka,Chatziioannou:2014coa,O'Shaughnessy:2014dka}.
However, since we only need an approximate error estimate that can be obtained
in a fast and reliable way across the \BBH parameter space, we choose to use
the Fisher-matrix method here for nonprecessing binaries, and we neglect small
correlations with extrinsic parameters such as sky location, orientation or
distance.

The characteristic standard
deviations
in the measurement process are estimated by \citep{Poisson:1995ef}
\begin{eqnarray}
 \Delta \theta^{i} &=& \sqrt{(\Gamma^{-1})_{ii}},  \label{eq:deltaTheta_Fisher}
\\
 \Gamma_{ij} &=& \left( \frac{\partial h}{\partial\theta^{i}}, \frac{\partial
h}{\partial
\theta^{j}} \right), \label{eq:Fisher_def}
\end{eqnarray}
where $\Gamma_{ij}$ is the Fisher information matrix and $h = h(\theta^i)$ is
the waveform model. The inner product used in \eqref{eq:Fisher_def} is given by
\begin{equation}
\left( g | h \right) = 4 \Re \int_{f_{\rm low}}^{\infty} \frac{\tilde{g}(f)
\tilde{h}^{*}(f)}{S_n (f)} df
\end{equation}
which is consistent with the \SNR definition in \eqref{eq:snr_def}. The form of
the waveform model we use allows us to calculate the partial derivatives used
in the definition of $\Gamma_{ij}$ analytically, and we ensure numerical errors
in the matrix inversion remain small.%
\footnote{In fact, we find that no element of
$\Gamma \Gamma^{-1}$ and $\Gamma^{-1}\Gamma$ deviates from the
respective element of the identity matrix by more than $10^{-7}$, in most
cases the deviation is much less than this.}

The only parameter we use to distinguish \BBH populations in this study is the
observable, redshifted chirp mass,
$\mathcal M^\ast$. The data sets of expected observable chirp
masses that we prepared following the algorithm introduced in
Sec.~\ref{subsec:rates} shall now be skewed further by adding measurement errors
to each binary in the data set. We do so by assuming a Gaussian distribution
centred around the chirp mass value of each binary with a standard deviation
given by the Fisher matrix estimate \eqref{eq:deltaTheta_Fisher}.  
\change{We evaluate
the Fisher matrix at the appropriate observed chirp mass and mass ratio of the 
binary, setting the value of the black hole spins to zero (although we allow the
spins to vary when calculating the Fisher matrix).  This has a negligible effect 
on our results as the measurement accuracy for the chirp mass is only weakly
dependent on the spins \citep{Ohme:2013nsa}.}
We randomly draw a sample from this distribution to re-define the measured chirp
mass. Similarly, when we later simulate the universe with a particular model,
each observation is drawn from the distribution that incorporates observational
biases, but the actually measured chirp mass is additionally offset following
the Gaussian distribution that simulates measurement errors.

\begin{figure}
	\centering
	\includegraphics[width=\linewidth]{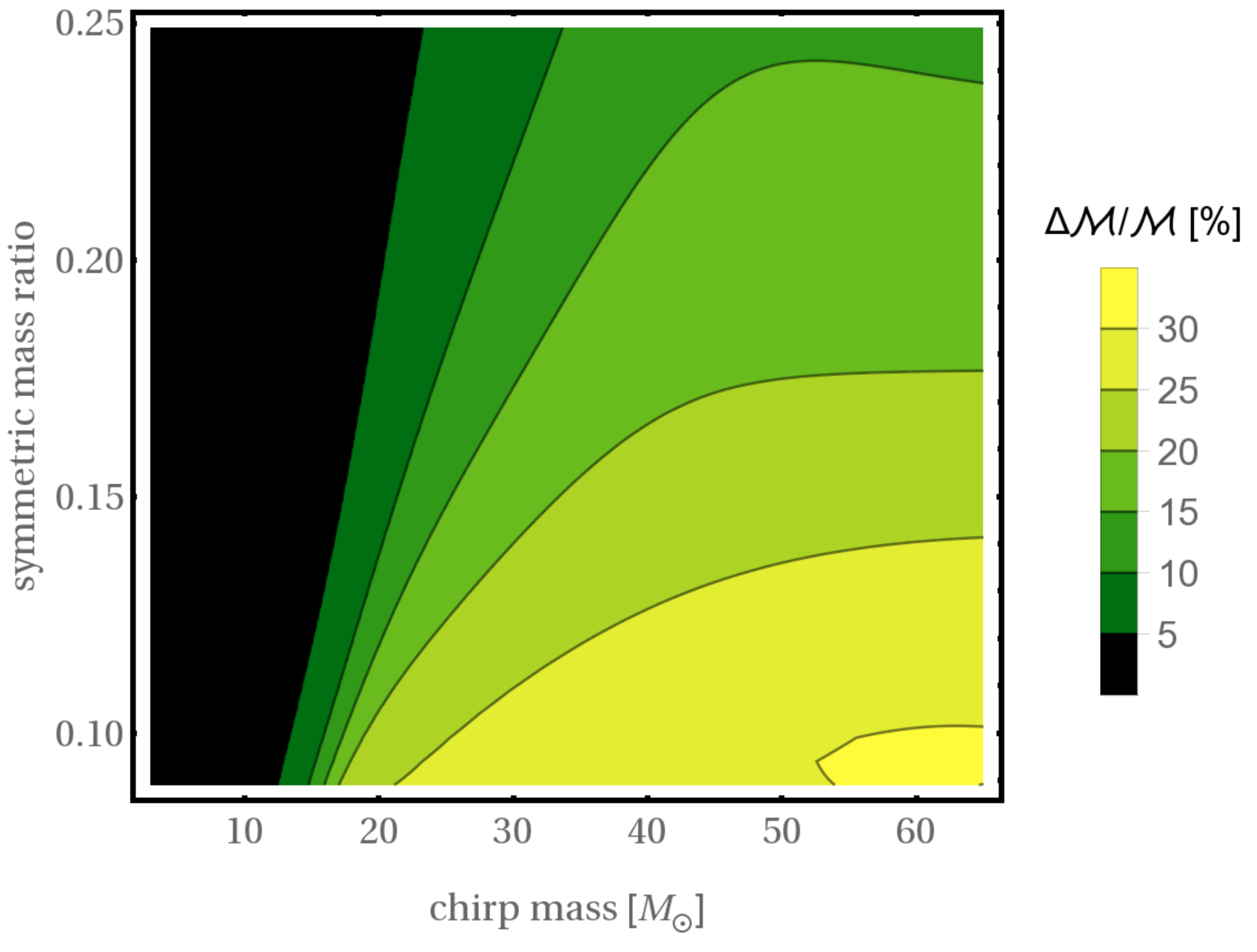}
	\caption{Expected relative measurement errors in the chirp mass for an
early configuration of \aLIGO, SNR 10, calculated using the PhenomC waveform
model \citep{Santamaria:2010yb}.}
	\label{fig:statErrMCPlot}
\end{figure}

The Fisher-matrix estimates scale inversely with the \SNR, so we only
calculate them once across the parameter space and
scale them for each binary in the data set according to its \SNR, which in turn
is inferred from the distance and a randomly chosen
orientation. Figure~\ref{fig:statErrMCPlot} shows the chirp mass uncertainty
at a constant \SNR of 10 across the parameter space for the early configuration
of \aLIGO.

\begin{figure*}[hp] 
	\centering
	\includegraphics[width=\textwidth]%
	{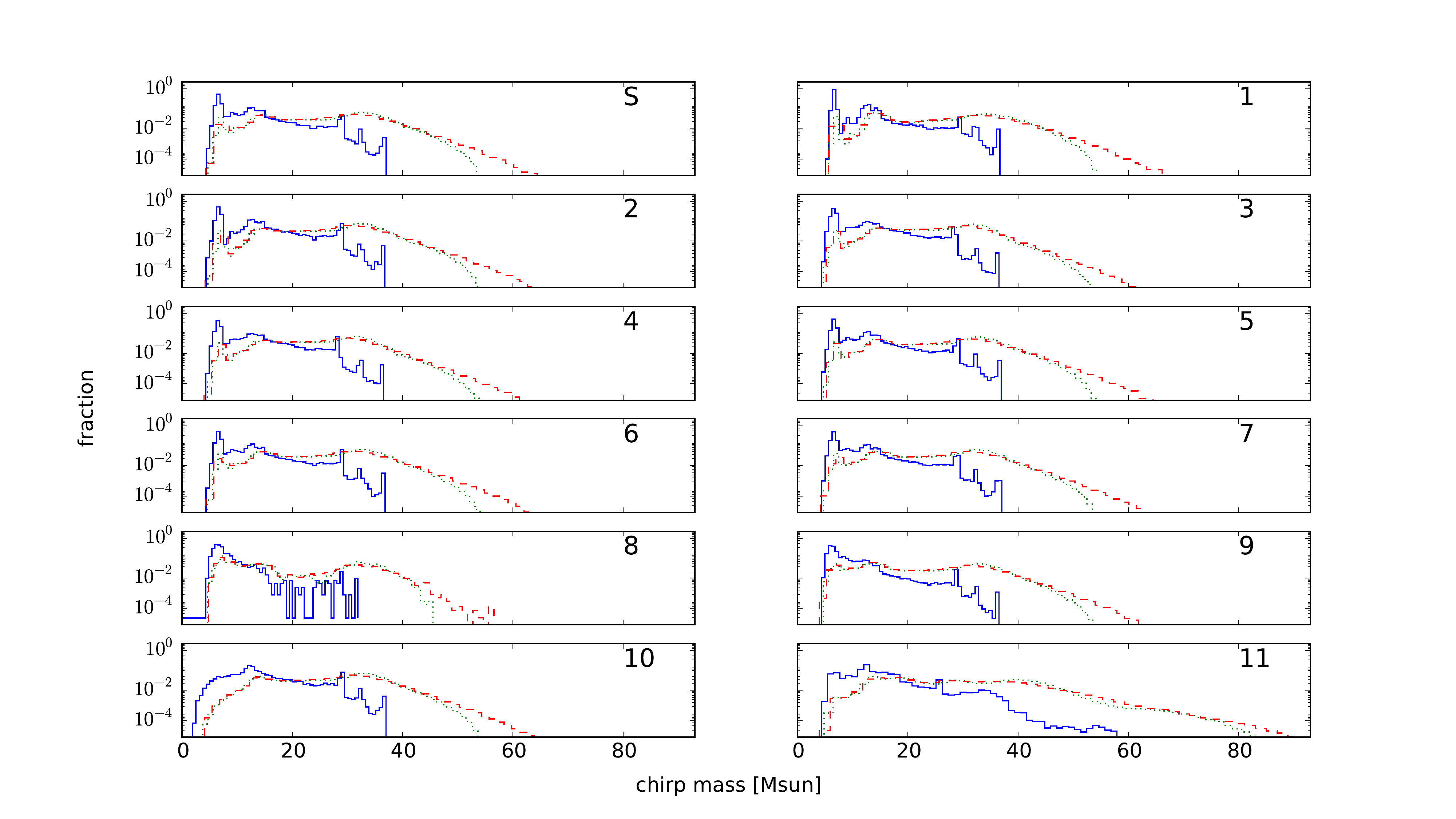} \vfill 
	\includegraphics[width=\textwidth]%
	{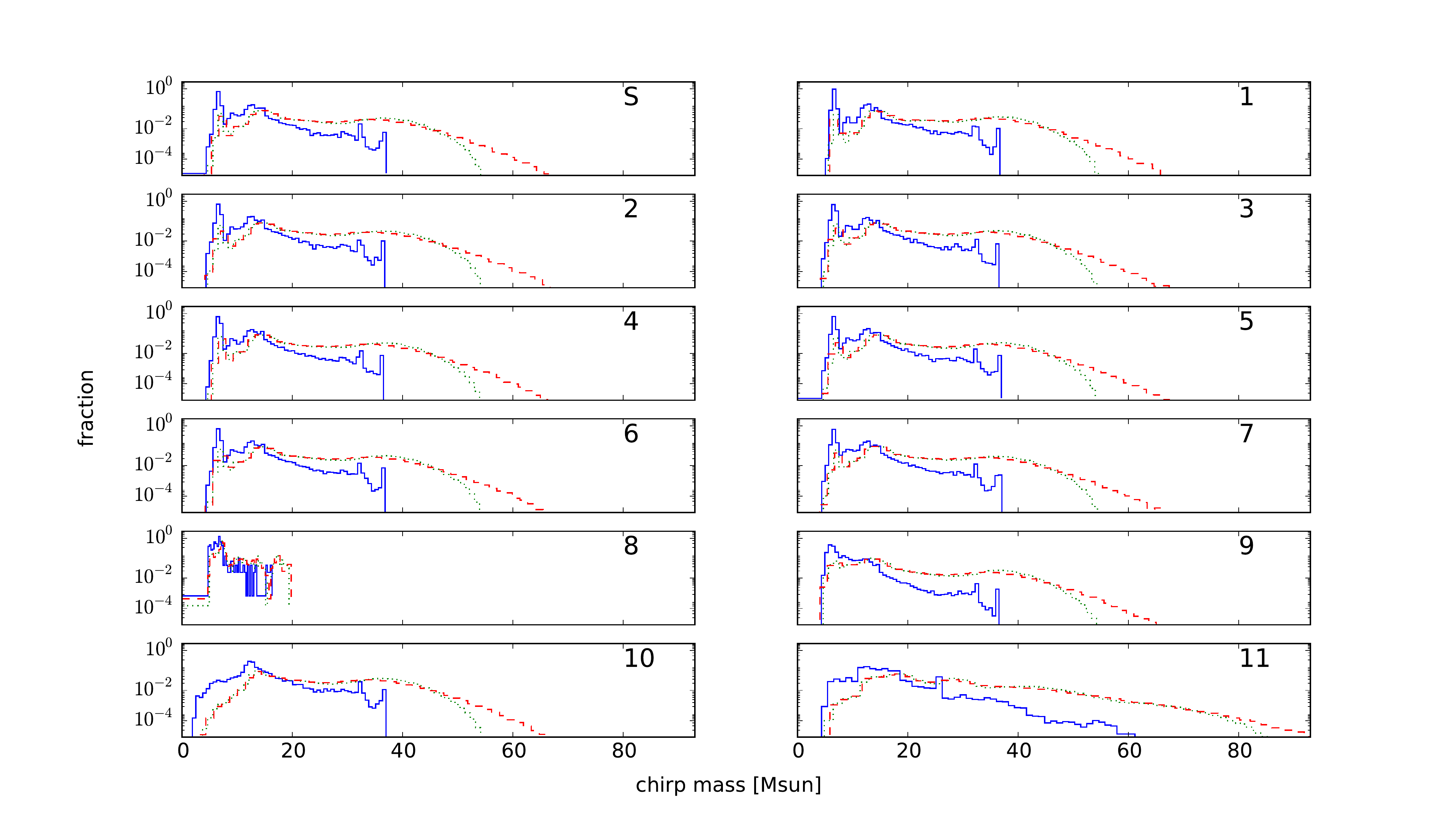}
	\caption{Chirp-mass distributions for each model in
\citet{Dominik:2012kk} using either their optimistic (top panel) or
pessimistic (bottom panel) submodel and a 50-50 split of solar and sub-solar
metallicities. The solid (blue online) line shows the intrinsic distribution as
given by \citet{Dominik:2012kk}. The dotted (green) line shows the same
distribution when accounting for the observational biases introduced in
Secton~\ref{subsec:rates} as predicted for the early configuration of \aLIGO.
Finally, the dashed (red) line with largest chirp-mass support shows the
expected observed distribution that additionally folds in the errors in
measuring the chirp masses through \GW observations.}
	\label{fig:skewedmassdistributions}
\end{figure*}

Figure~\ref{fig:skewedmassdistributions} illustrates the transition from the
intrinsic \BBH population, predicted by \citet{Dominik:2012kk} for each of
their models, to the expected observed chirp mass distribution. The main effect
of the observational bias detailed in Sec.~\ref{subsec:rates} is that the
distribution becomes skewed towards high-mass binaries, and its support
extends to larger chirp masses due to the redshift of distant sources. The
addition of measurement errors hardly affects the distribution at low
chirp masses, simply because the errors are small compared to the typical
variation of the distribution in this regime. For heavy systems, on the other
hand, noise fluctuations introduce a non-vanishing chance of measuring chirp
mass values greater than the largest (redshifted) chirp mass in each model.
Hence, the main effect of introducing measurement errors is that the
expected observed distributions show a characteristic tail at high chirp masses.

\section{Combining Measured Rates and Chirp Masses}
\label{sec:calculating_probability}

Given a set of \BBH observations, for each model variation $V_i$, 
we wish to calculate the posterior probability for that being the correct
model. The information we gather about the correct model is twofold. First, we
obtain a set of observed chirp masses $\{ \mathcal M\}$, and second, we 
measure the rate of \BBH detections by observing $n$ binaries in a given
observation period.  
\change{In reality each observation will include measurements of additional physical parameters of the 
system, such as the component spins (see \cite{PhysRevD.87.104028,PhysRevD.89.124025} for 
information on how measurements of spin misalignments can help to constrain astrophysical 
formation scenarios.) Including additional information from these other dimensions should help in 
distinguishing astrophysical formation scenarios.

We simulate the observed population} by choosing one of the model variations, adjusted to
account for selection effects as described above, to describe the universe. We
then draw $n$ individual chirp mass measurements from this model, which form
the data $\{\mathcal M\}$. The number of observations we assume is itself
drawn from a Poisson distribution with a mean value that is dictated by the
observation time and the merger rate of the model variation we have selected to
simulate the universe.

With these measurements, $\{ \mathcal M \}$ and $n$, the posterior probability
that model $V_i$ is the correct model reads
\begin{equation}
P(V_i \vert \{\mathcal M\}  , n) = \frac{P(\{\mathcal M\} , n | V_i)
P(V_i)}{P(\{\mathcal M\} , n)},
\end{equation}
where we have used Bayes' Theorem. $P(V_i)$ is the
prior probability on model $V_i$, $P(\{\mathcal M\} , n \vert V_i)$ is the
likelihood of making these
chirp mass measurements \emph{and} measuring this detection rate given 
model variation $V_i$, and $P(\{\mathcal M\} , n)$ is a normalisation factor
called the evidence. 

Assuming that the number of observations $n$ is independent from the chirp mass
values we observe, we can rewrite this as
\begin{equation}
P(V_i | \{\mathcal M\}  ,n) = \frac{P( \{\mathcal M\} | n, V_i) P( n | V_i)
P(V_i)}{P(\{\mathcal M\} , n)}.
\end{equation}
We normalise by assuming that the discrete model variations we consider cover
all possible states of the universe, which is an idealisation that we shall
discuss in more detail later. However, this assumption allows us to define
the normalisation factor by requiring the sum of the probabilities for
each model to be unity, which leads to 
\begin{equation}
P(V_i | \{M\}  , n) = \frac{P( \{M\} | n, V_i) P( n | V_i) P(V_i)}{ \sum_k P(
\{M\} |n, V_k) P( n | V_k)
P(V_k)}.
\end{equation}

We assume a uniform prior on the models,
\begin{equation}
P(V_i) = \frac{1}{\mathcal N},
\end{equation}
where $\mathcal N$ is the number of models we are considering. The prior
then cancels out and we are left with
\begin{equation}
P(V_i | \{\mathcal M\} , n) = \frac{ P( \{\mathcal M\} | n, V_i) P( n | V_i)}{
\sum_k P( \{\mathcal M\} | n, V_k) P( n | V_k)}.
\label{eq:probab_noprior}
\end{equation}

The likelihood of making $n$ observations in a set time, given a model
predicting mean number of observations, $\mu_i$, is given by the
Poisson distribution:
\begin{equation}
P(n | V_i) = P(n | \mu_i) = \frac{e^{-\mu_i} \mu_i^{n}}{n!}. 
\label{eq:poisson}
\end{equation}

The likelihood terms of the form $P(\{\mathcal M\} | n,  V_i)$ are calculated by
binning the chirp mass distributions for each model into a histogram. We then
calculate the likelihood of the observed samples being drawn from their bins 
using the multinomial distribution
\begin{equation}
P(\{\mathcal M\} | n, V_i) = n! \prod_{k=1}^{b} \frac{p_{ik}^{x_k}}{x_k !},
\label{eq:multinomialdistribution}
\end{equation}
where $n$ is the number of samples in the observations, $b$ is the number of 
bins, $p_{ik}$ is the probability in model $i$ of drawing a sample from
bin $k$ and $x_k$ is the
number of observations that fall into bin $k$, with
\begin{equation}
\sum_k x_k = n \quad \textrm{and} \quad \sum_k p_{ik} = 1.
\end{equation}

We calculate $p_{ik}$ for each model and bin as the fraction of the total number
of samples in the model which fall into that bin. The bin size we employ is
motived by Scott's rule \citep{SCOTT01121979},
\begin{equation}
\Delta = \frac{3.5 \sigma}{\sqrt[3]{N_m}}, \label{eq:Scotts_binwidth}
\end{equation}
where $\Delta$ denotes the bin width, $\sigma$ is the standard deviation of the
model, and $N_m$ is the total number of samples in model. To be able to
consistently compare our simulated data with all models, we apply
(\ref{eq:Scotts_binwidth}) to all models and then use the median bin width for
the actual analysis. However, we find that changing this bin width by a factor
of a few does not impact our results noticeably.

\section{Observing Scenarios}
\label{sec:observing_scenarios}

The method we have developed transforms predicted binary distributions and
merger rates into \emph{observable} distributions and detection rates which in
turn can be confronted with a set of observations in order to assign posterior
probabilities to each model. As such, the method is generally applicable to any
set of theoretical predictions and detector configuration.

In the following, however, we present results for specific choices of binary
population models, detector sensitivity and observing time. As detailed
Sec.~\ref{subsec:models} and summarised in Table~\ref{tab:models}, we consider
12 binary population models by \citet{Dominik:2012kk}, each with both the
``pessimistic'' (submodel B) and ``optimistic'' (submodel A) assumption
about the common envelope evolution. This leads to 24 distinct predictions
of the \BBH chirp mass distribution (see
Figure~\ref{fig:skewedmassdistributions}), where each comes with a distinct
average merger rate density that we take from the arithmetic mean of the solar
and subsolar metallicity predictions by \citet{Dominik:2012kk} (Tables 2 and 3
therein). The local merger rate
densities for each model are given in Table~\ref{tab:expectedrates}.
Interestingly, due to the mass-dependent observational bias, models with higher
merger rate density do not necessarily exhibit a higher detection rate, see for
instance models 9 and 10 in Table~\ref{tab:expectedrates}.

Recent calculations by \citet{2013ApJ...779...72D} that
include the cosmological evolution of merger rates give lower rate densities
than the ones we infer from earlier work of the same authors. Consequently, the
detection rates we find are up to a factor of 2 larger than those recently
predicted by \citet{2014arXiv1405.7016D} (this is based on a direct comparison
of our method with their otherwise equivalent approach using the same detector
configuration). However, this neither affects the general proof of principle
carried out here, nor do the conclusions we shall draw in the following section
change qualitatively by varying the detection rate at this level. 

\begin{deluxetable}{rrrrrrrrr}
 \tablecolumns{9}
 \tablecaption{Predicted merger and detection rates.
\label{tab:expectedrates}}
\tablehead{
 &  \multicolumn{2}{c}{$\bm{\mathcal R}$\tablenotemark{a}} &
 \multicolumn{2}{c}{$\bm{\langle \mathcal{M} \rangle}$\tablenotemark{b}} &
\multicolumn{2}{c}{$\bm \mu$\tablenotemark{c} \textbf{(O1)}} &
\multicolumn{2}{c}{$\bm
\mu$\tablenotemark{c} \textbf{(O2)}} \\
$\bm{V_i}$ & 
\colhead{B} & \colhead{A} 
& \colhead{B} & \colhead{A} &
\colhead{\hspace{9pt}B} & \colhead{A} &
\colhead{\hspace{9pt}B} & \colhead{A}
} 
\startdata
 0 & 7.8 & 40.8 & 26.0 & 24.9 & 4.0 & 25.2 & 64 & 402 \\ 
  1 & 4.6 & 6.8 & 27.3 & 26.2 & 2.3 &3.9 & 37 & 63 \\ 
  2 & 8.3 & 36.0 & 26.6 & 24.9 & 4.2 & 25.9 & 67 & 413 \\
  3 & 4.0 & 47.6 & 25.0 & 24.4 & 1.9 & 28.7 & 30 & 458 \\ 
  4 & 0.1 & 3.1  & 25.0 & 24.7 & 0.1 & 1.9 & 1 & 30 \\
  5 & 7.8 & 40.9 & 26.0 & 24.9 & 4.0 &25.3 & 64 & 404 \\ 
  6 & 7.9 & 41.3 & 25.6 & 24.2 & 3.9 &  25.1 & 63 & 401 \\
  7 & 8.6 & 47.1 & 25.3 & 23.8 & 4.0 &26.3  & 65 & 420 \\ 
  8 & 0.4 & 2.1  & 21.3 & 10.0 & 0.0  &0.6  & 1 & 9 \\
  9 & 11.8 & 54.6 & 23.2 & 20.7 & 3.4 &20.2 & 54 & 324 \\ 
 10 & 5.8 & 31.3 & 26.8 & 26.2 & 4.3 & 26.0 & 68 & 415 \\%
11& 10.4 & 54.5 & 29.8 & 28.6 & 8.5 & 46.5 & 136 & 742 
\enddata
\tablenotetext{a}{Local merger rate density in $\textrm{MWEG}^{-1}
\textrm{Myr}^{-1}$.}
\tablenotetext{b}{Average \change{predicted} observed chirp mass in \Msun (see Sec.~\ref{sec:observationalbias})}
\tablenotetext{c}{Mean number of detections predicted by each model for the
early \aLIGO observing runs O1 and O2 (see text for details).}
\tablecomments{The binary populations models, $V_i$, predicted by
\citet{Dominik:2012kk} are summarised in Table~\ref{tab:models} and the
submodels B and A refer to pessimistic and optimistic assumptions about the
common envelope evolution of Hertzsprung gap donors (Sec.~\ref{subsubsec:hg}).}
\end{deluxetable}

We also have to specify in the sensitivity
(i.e., noise spectral density) of our assumed \GW detector and the observing
time. Closely following \citet{Aasi:2013wya}, we consider the first two \aLIGO
science runs dubbed O1 and O2, respectively. The first science run for \aLIGO
(O1) is planned to begin in autumn 2015. The duration of O1 will be
approximately 3 months for the two \aLIGO detectors. We assume each detector
has a duty cycle of 0.8 so that the total period of coincident observation
during O1 will be about 0.16 years. The noise power spectral density we use is
the ``early \aLIGO'' configuration \citep{Shoemaker:2010}.

We further consider a second science run, O2. During O2, the detectors are
planned to observe for approximately 6 months with a comparable duty cycle to
O1. It is expected that, after further improvements of the instruments following
O1, the \aLIGO detectors during O2 will be approximately a factor of 2 more
sensitive than the nominal early \aLIGO noise curve we use for O1. While the
evolution of the noise power spectral density is in general a function of the
frequency, we find that, in practice, the difference between the predicted noise
curves in \citet{Aasi:2013wya} results in improved horizon distances and error
estimates that are well approximated by simply scaling the results we obtain for
the early \aLIGO configuration. Hence, we simulate O2 by multiplying the O1
horizon distance by 2. The Fisher-matrix errors change only due increased \SNR
at fixed distance. This increase in sensitivity leads to a factor of 8 increase
in volume meaning that, in total, O2 surveys 16 times the time-volume of O1. 
We show in the following section that this is when we will begin to
distinguish between astrophysical models.

\section{Results: Distinguishing \BBH Formation Models}
\label{sec:results}

\subsection{First \aLIGO observing run (O1)}
\label{subsec:O1}

\begin{figure}
	\centering
	\includegraphics[width=0.48\textwidth]{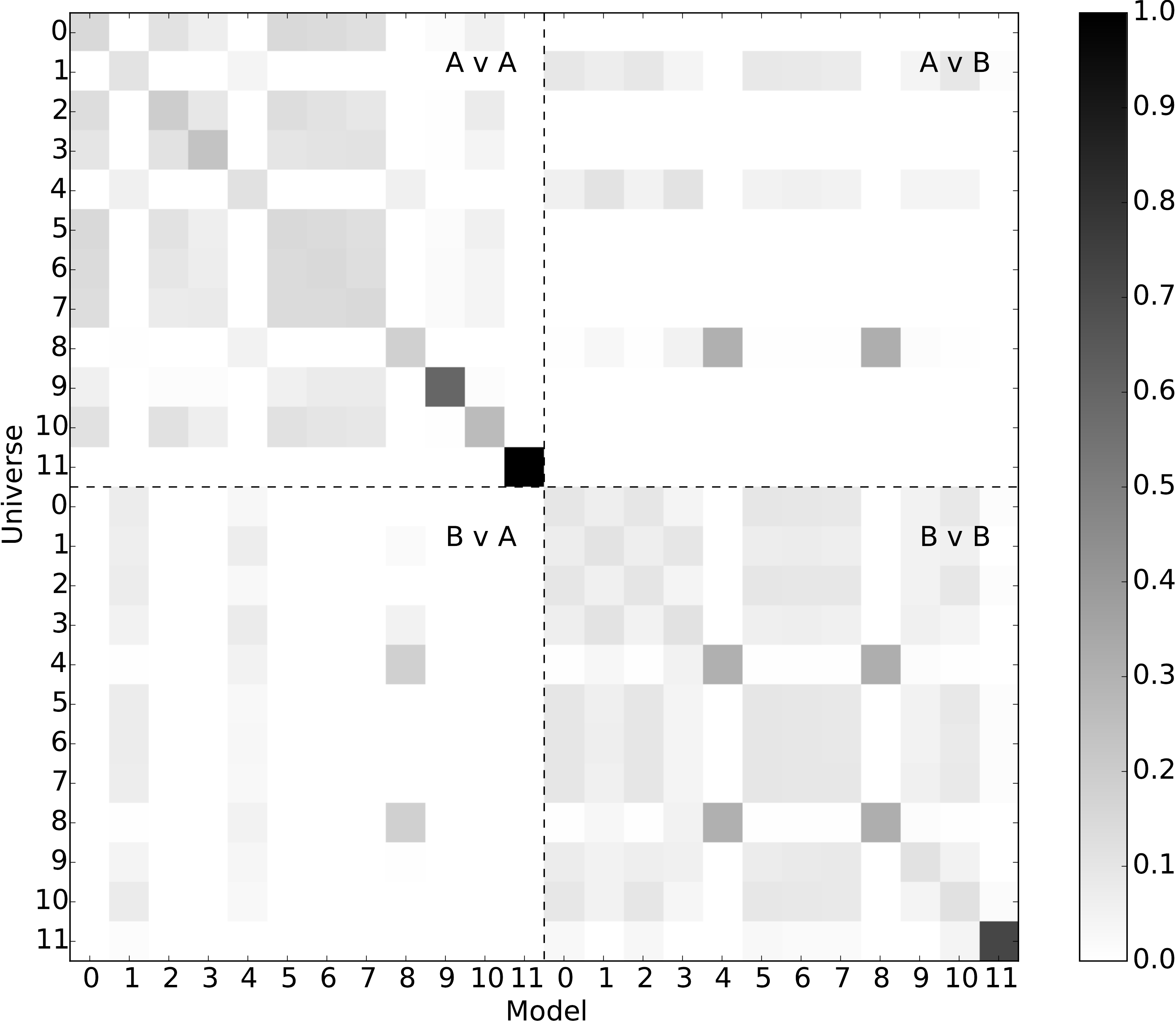}
	\caption{The median posterior probability for each model in the
set of \dominik models after an O1 like observing period of 0.16 years, calculated from 10000 repeats. The
model which observations were drawn from is shown on the axis labelled
\emph{Universe}. The models which these observations were then compared to is
labelled \emph{Model}, 
\change{so that the probabilities in each row sum to one.}
Models 0-11 are described in Table 1. The two submodels,
A and B, are described in Section~\ref{subsubsec:hg}.}
	\label{fig:01matrixplot}
\end{figure}

\change{We simulate the observed \BBH systems,
assuming the universe matches one of the models from
\citet{Dominik:2012kk}, and calculate the posterior probability for
each model.  We repeat the
experiment 10000 times before turning to the next model to simulate the
universe. Figure~\ref{fig:01matrixplot} gives the median posterior
probability for each model.  

In cases where one or
few models have a high probability, these would be distinguishable
from the other models.  However, all models with a high probability
would be consistent with the observations. We reiterate that here we
restrict attention to the models in \cite{Dominik:2012kk}.  Of course 
these do not cover the full space of binary merger predictions.  If we
were to include a broader range of models, it is likely that the conclusions we
are able to draw would be weaker as various models would lead to comparable
rates and mass distributions.  Nonetheless, some of the conclusions we reach,
such as excluding a number of models if there are no observations in O1, are
robust.}

We first observe that, for the most part, we would be able to distinguish
between submodels
A and B that correspond to different common envelope scenarios (see
Sec.~\ref{subsubsec:hg}). This is unsurprising as the predicted
rates for the majority of models are significantly higher for submodel A (cf.\
Table~\ref{tab:expectedrates}).
Models which predict low detection rates for model A remain degenerate with those in model B.
The mass distribution from such a small sample does not provide enough
additional information to break these degeneracies in the rates. For example,
model 1 A uses a very high, fixed envelope binding energy, meaning
that most binaries entering a common envelope event fail to throw off the common
envelope and merge, causing them to never form \BBH systems (for a
more detailed discussion of this, see \citet{Dominik:2012kk}). On the other
hand, submodel B does not allow a binary to survive a common
envelope event if the donor is on the Hertzsprung Gap, and so again, many
binaries merge and never form \acp{BBH}. This generically lowers the
merger rates and thus detection rates for submodel B models, leading to the
degeneracy visible in the upper right quadrant of Fig.~\ref{fig:01matrixplot}.

Another interesting example involves models 4 and 8 that, in the pessimistic
submodel B, are consistent with no observations at all during O1. Hence, they
cannot be distinguished from each other, or indeed model 8 A, although they are
favoured over all other models if indeed no detection are made.

Within the two submodels, it is difficult to identify the correct model. Indeed,
there are numerous variations which would be indistinguishable from the standard
model.  The only model which can be clearly identified is model 11, a model
which reduces the strength of stellar winds by a factor of 2 over the standard
model. We now discuss why we are able to distinguish this model from the others
in such a short observational period. 

\subsection{Stellar winds}
\label{subsubsec:stellarwinds}

In massive O-type stars, stellar winds of high temperature charged gas are
driven by radiation pressure. In Wolf--Rayet stars mass loss rates can be as
high as $10^{-4} \Msun \text{yr}^{-1}$ \citep{2002A&A...389..162N}. This can cause stars to lose
a large amount of mass prior to the supernova. Theoretical uncertainties in
modelling these mass loss rates therefore translate into uncertainties in the
pre-supernova masses for massive stars. \citet{Dominik:2012kk} examine the
effects of reducing the strength of stellar winds by a factor of 2 on the
distribution of \acp{BBH} in their Variation 11. Firstly, reducing stellar winds
results in stars having a higher mass prior to supernova than they would
otherwise have. This in turn leads to more mass falling back onto the compact
object during formation, which reduces the magnitude of natal kicks given to
black holes. This results in more systems surviving the supernova (rather than being
disrupted) and increases the merger rates. More massive
pre-supernova stars also form more massive remnants, resulting in the most
massive \BBH having a chirp mass of $\sim64 \Msun$ with reduced stellar winds
compared to $\sim37 \Msun$ using the standard prescription. Finally, reducing
the strength of stellar winds allows stars with a lower zero age main sequence
mass to form black holes due to more mass being retained. This can boost the \BBH merger rate compared to the standard model.

\begin{figure}[ht]
	\centering
	\includegraphics[width=0.5\textwidth]%
{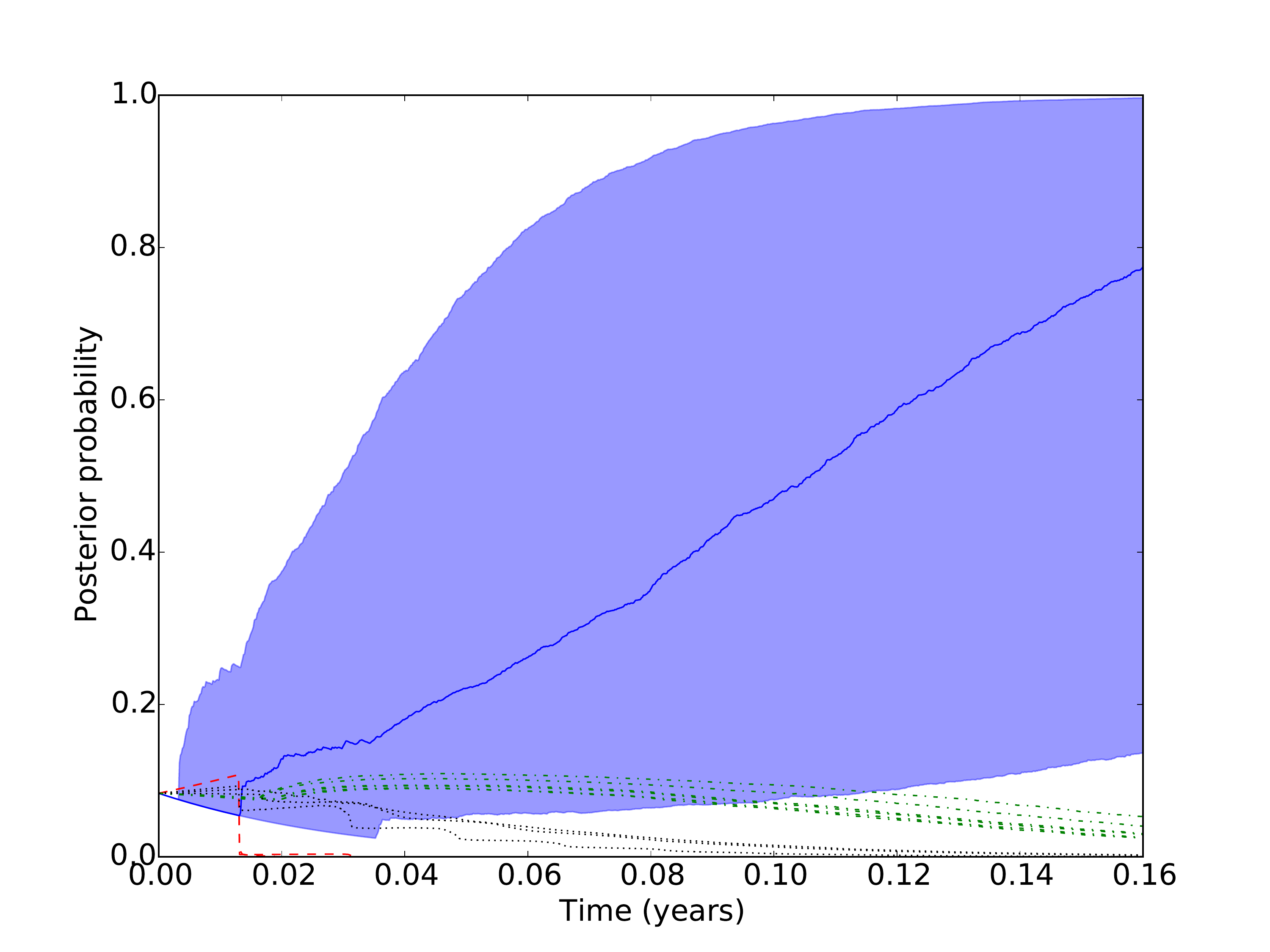}
	\caption{The median posterior probability for each of the models in the set as a function of observation time for a period of time corresponding to the \aLIGO O1 run (0.16 years). \GW observations are drawn from a universe following Variation 11, submodel B which reduces the strength of stellar winds by a factor of 2 compared to the standard model. The blue (solid) line shows the median posterior probability for Variation 11 taken from 10000 repeats, and the shaded error bar shows the 68\% confidence interval. Variations 0,2,5,6,7 \& 10 are plotted in green (dot-dash), while variations 1,3 \& 9 are plotted in black (dotted). Variations 4 \& 8 predicting $\sim 0$ observations in O1 are plotted in red (dashed).}
	\label{fig:sampling11functime}
\end{figure}

All of these effects combined mean that Variation 11 predicts \acp{BBH} with characteristically higher chirp masses, as well as predicting a much higher merger rate than all other models (even for the pessimistic submodel B in O1, Variation 11 predicts $\mathcal{O}(10)$ observations). We therefore expect that we would be able to correctly distinguish a universe following Variation 11 from all other models with relatively few observations. In Figure~\ref{fig:sampling11functime} we show the median posterior probability for each model as a function of the observation time, based on 10000 redraws of the observations. We find that when drawing observations from a universe following Variation 11 we overwhelmingly favour it within the duration of O1, with $\mathcal{O}(10)$ observations. 

\subsection{Second \aLIGO observing run (O2)}
\label{subsec:O2}

\begin{figure}
	\centering
	\includegraphics[width=0.48\textwidth]{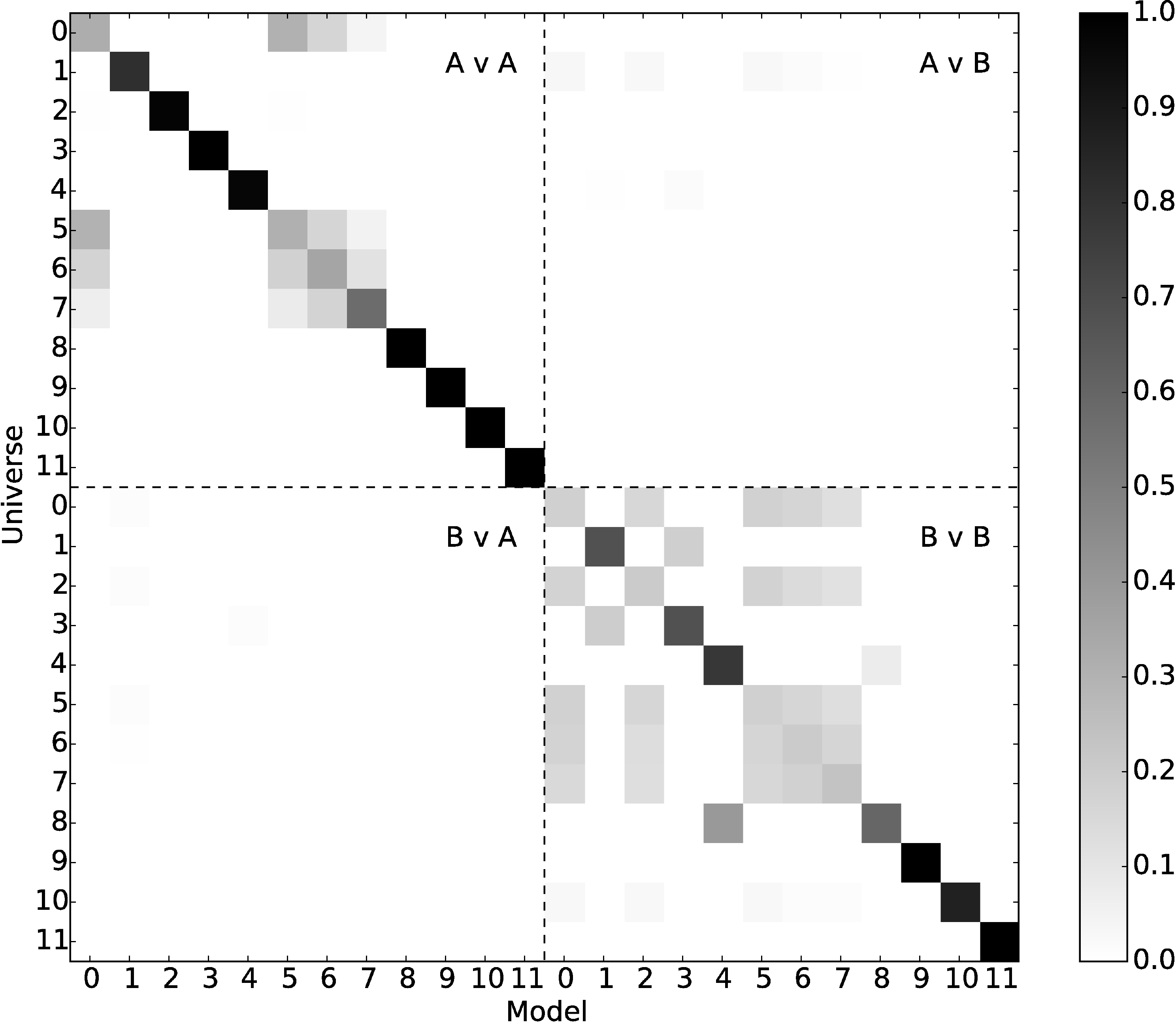}
\caption{The median posterior probability for each model in the
set of \dominik models after an O2 like observing period of 0.32 years with
a detector more sensitive than the early \aLIGO noise curve by a factor of 2. The median is calculated based on 10000 redraws of the observations. The model which observations were drawn from is shown on the axis labelled \emph{Universe}. The model which these observations were then compared to is
labelled \emph{Model}. Models 0-11 are described in Table~\ref{tab:models}. The
two submodels, A and B, are described in Section~\ref{subsubsec:hg}.}
	\label{fig:02matrixplot}
\end{figure}
 
We now turn our attention to the second observing run, O2, and investigate
which models can be distinguished using the much larger
time-volume surveyed by O2. In Figure~\ref{fig:02matrixplot} we again show a
matrix plot showing the (median) posterior probability for each model after a
period corresponding to the O2 run. 

Figure~\ref{fig:02matrixplot}
has a more diagonal form than Figure~\ref{fig:01matrixplot},
meaning that in many cases the correct model is favoured and 
others are disfavoured within the O2 period. In particular, the optimistic and
pessimistic submodels A and B become almost entirely distinct from each other. 
This is because most of the \dominik models predict $\mathcal{O}(100)$ 
($\mathcal{O}(10)$) observations during the O2 period for the optimistic
(pessimistic) submodels respectively (as shown in
Table~\ref{tab:expectedrates}). Furthermore, the majority of variations in 
submodel A can be unambiguously identified; the exception being that 
the standard model which remains degenerate with models 5, 6 and 7, as we discuss
in detail in Section \ref{subsubsec:kicks}.   For the pessimistic submodel B,
the standard model
remains indistinguishable from a number of other variations.  However, there are
a few models which can be clearly distinguished, including models 4 and 8 (that predict
significantly lower rates), and 9, 10 and 11.  All of these models predict tens of
observations and consequently, we are able to use information
from both the chirp mass distribution and the detection rate to help distinguish
models.  Model 10 involves the variation of the supernova engine, which we elaborate
on in Section \ref{subsubsec:supernova}.

\subsubsection{Black hole kicks and maximum neutron star mass}
\label{subsubsec:kicks}

Not all models are distinguishable, even with the $\mathcal{O}(100)$ observations predicted by the optimistic submodel A for O2. For example, in Figure~\ref{fig:02matrixplot} we see that the standard model is degenerate with Variations 5, 6 and 7. We now explain why this is so.

As already mentioned, it is unclear what the correct distribution of natal kicks
given to black holes upon formation is. In order to investigate the
possibilities, \citet{Dominik:2012kk} vary two parameters relating to the kicks
imparted onto newly formed black holes; the characteristic velocity $\sigma$ and
the fraction of mass $f_b$ which falls back onto the newly born black hole.

In their standard model, black holes receive a kick $v_k$ whose magnitude $v_\text{max}$ is drawn from a Maxwellian distribution with $\sigma = 265 \text{km s}^{-1}$, and reduced by the fraction of mass falling back onto the black hole $f_b$ as
\begin{equation}
v_k = v_\text{max} (1 - f_b),
\end{equation}
where $f_b$ is calculated using the prescription given in \cite{Fryer:2011cx}.

In order to test the effects of smaller natal kicks, in Variation 7 \dominik
reduce the magnitude of kicks given to neutron stars and black holes at birth by
a factor of 2. They use a Maxwellian distribution with $\sigma = 132.5 \text{km
s}^{-1}$. For \acp{BBH}, this has very little effect on the chirp mass
distribution, and so one cannot expect to be able to distinguish this model
from one using full kicks. 

The same holds true when the maximum
neutron-star mass is increased (decreased) from its fiducial value in the
standard model of $2.5 \Msun$. This has very little impact on the \BBH chirp
mass distribution and so there is effectively a degeneracy between these models.
This could be resolved by also including \BNS observations in the comparison. We
do not do this here as we concentrate on the \BBH predictions, due to the
prediction by \citet{Dominik:2012kk} that these will dominate the early \aLIGO
detections.

\subsubsection{Supernova engine}
\label{subsubsec:supernova}

In their standard model, \dominik employ the \cite{Fryer:2011cx} prescription to
calculate the fraction of mass falling back onto the black hole during
formation, and thus the black hole masses. In particular, they use the
\emph{rapid} supernova engine. When employed in a compact binary population code
such as \texttt{StarTrack}, the rapid supernova engine reproduces the observed
mass gap \citep{Ozel:2010su,2011ApJ...741..103F} in compact objects between the
highest mass neutron stars and the lowest mass black holes (for a discussion of
using \GW observations to infer the presence or absence of a mass gap, see
\cite{Hannam:2013uu,2015arXiv150303179L,2015MNRAS.450L..85M}).

\change{In model 10 \dominik vary this prescription to use the \emph{delayed} supernova
engine from \cite{Fryer:2011cx}, which produces a continuous distribution of
black hole masses (and thus \BBH chirp masses). We therefore expect that the
difference between these two models might be visible in the chirp mass
distributions. We see however from Table~\ref{tab:expectedrates} that these two
models predict similar merger rates for \BBH, and so we do not expect to be able
to distinguish them based on the detection rate. Nonetheless, we see from 
Figure~\ref{fig:02matrixplot} that this model
can be distinguished from the others by the end of O2 and
even, to a lesser degree, at the end of O1 (Figure~\ref{fig:01matrixplot})}

\begin{figure*}
	
\begin{tabular}{cc}
\textbf{Rates only} & \textbf{Mass distribution only} \\
 \includegraphics[width=0.48\textwidth]{%
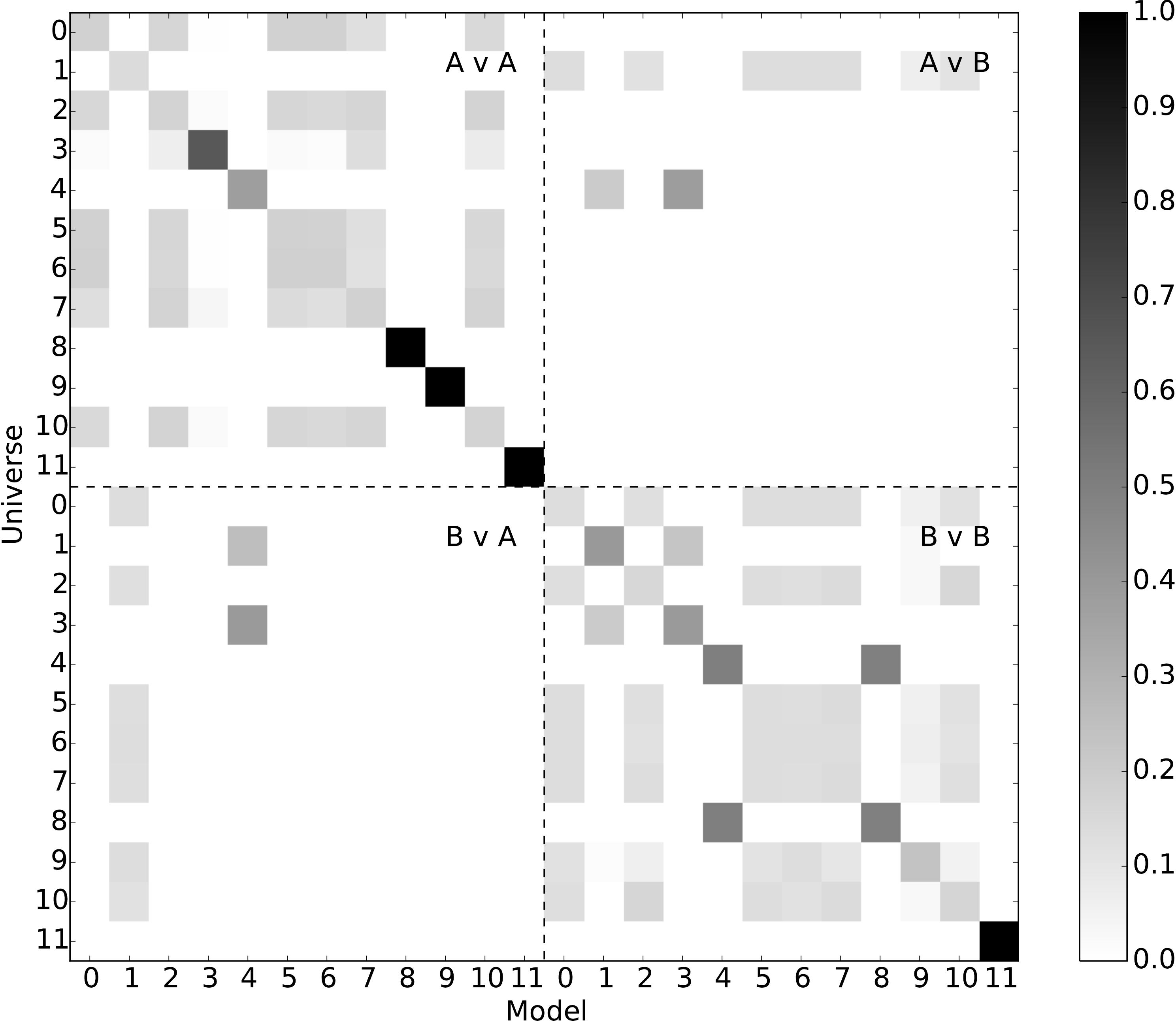}
\hspace{10pt} & \hspace{10pt}
\includegraphics[width=0.48\textwidth]{%
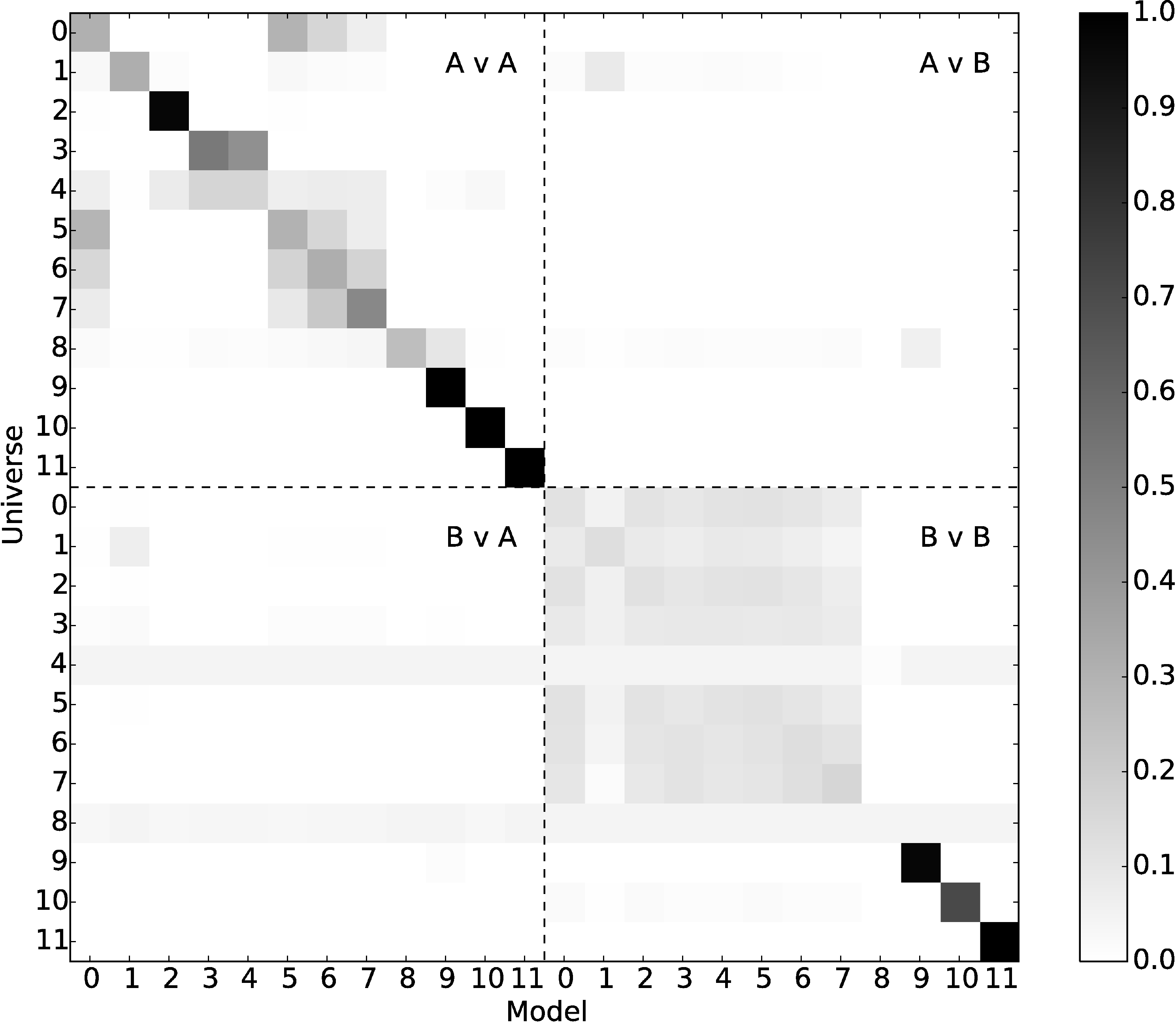}
\end{tabular}
	\caption{Probabilities for the scenario of
Fig.~\ref{fig:02matrixplot}, separated into contributions from the rates (left)
and the mass distribution (right).}
	\label{fig:matrixplotO2individual}
\end{figure*}

\change{To illustrate the importance of both the mass distribution and predicted rates,
in Figure~\ref{fig:matrixplotO2individual} we show the results that would be
obtained using only one of these to separate the models.  By comparing these
results with Figure~\ref{fig:02matrixplot}, it becomes clear that \textit{both} the
mass and rate measurements contribute significantly to our ability to distinguish
between models.  As expected, the delayed supernova engine (model 10) is 
distinguished from observed masses, but the rates are quite degenerate with
other models.  In contrast, models 4 and 8, are distinguished primarily by
rate measurements, and not masses.
As we have mentioned previously, the unknown spin distribution of black
holes in binary systems can change the rate by a factor of two or three.  Similarly,
both the mass and rate distributions are subject to uncertainties due to additional 
physical effects which are not yet incorporated.  Consequently, one might choose
to incorporate an uncertainty in the rates or mass distributions.  The results in Figure~\ref{fig:matrixplotO2individual} illustrate the extreme scenario where one assumes 
knowledge of only the rate or mass distribution.  Adding an uncertainty to the mass
or rate distributions will lead to a result between those shown in 
Figure~\ref{fig:02matrixplot} and \ref{fig:matrixplotO2individual}.
}



\section{Summary and future work}
\label{sec:conclusion}

In this paper we have outlined a method for comparing \GW observations of \BBH
mergers to binary population synthesis predictions using a Bayesian model
comparison framework. Starting from chirp mass distribution predicted by
\dominik, we produce predicted observed chirp mass distributions accounting for
known observational effects. We incorporate
\begin{enumerate}[(a)]
\item The redshifting of observed binary masses due to the cosmological distances out to which they will be observed.
\item The observational bias of \GW detectors to detect more massive systems,
since they can be seen to greater distances and thus in much larger volumes.
\item Fisher matrix estimates of measurement uncertainties in the recovery of
the chirp mass of \BBH.
\end{enumerate}

We show that given the merger rates predicted by the models of \dominik, 
we will begin to be able to distinguish between population synthesis models 
in the first two \aLIGO science runs. Ruling out models in turn can help to 
constrain the value of unknown parameters, which relate to poorly understood 
astrophysics relating to binary evolution.  

\change{
Of course, the set of models considered
here by no means encompasses the full set of stellar evolution models
available in the literature.  We restricted attention to this subset of models
as the data was publicly available in an easy to use form.  It would be
straightforward to include additional models into this analysis.
Ideally, we would make use of a dense set of models, where numerous
astrophysical parameters are jointly varied.  This would allow us to interpolate
between models, and extract best-fit parameters
 \citep{PhysRevD.88.084061,O'Shaughnessy:2006wh}.  
Furthermore, we have restricted attention to the two best-measured
quantities: the rate and chirp mass distribution of binaries, and only used
point estimates of the masses.  The inclusion
of full parameter distributions can only enhance our ability to distinguish
between models.}

\change{
The method we have introduced allows us to distinguish between a given set
of stellar evolution models.  It will identify the model, or models, that best
agree with the observed rate and mass distribution.  It will not, however, 
indicate whether the best model is actually a good fit to the observations ---
only that it is better than the others.  This could be remedied by introducing
a simple, generic model.  For example, the intrinsic mass distributions
shown in Figure~\ref{fig:skewedmassdistributions} are reasonably well 
desribed by a decaying power law with an upper and lower mass cutoff.
One could then imagine extending the set of models to include this 
phenomenological mass distribution parametrized by three variables
with an additional variable rate.  To calculate the posterior for this
distribution, we would then have to marginalize over four parameters.
Thus, even if the generic model was a reasonable fit to the data, it
would be penalized by the large initial parameter space.  It is likely
that the generic model would be preferred after a small number of
observations.  With a large number of observations, the rate and
mass distributions would be reasonably well measured.  Any specific 
model which matched the observations well would then be preferred
to the generic model due to its broader support on the parameter space.
It would be reasonably straightforward to extend our method to 
include a generic model, and this is something we plan to incorporate in
the future.
}

In this study we concentrated on the information that could be gained from \GW
observations of \BBH mergers. \aLIGO and \AdV are also expected to
observe the inspiral, merger and ringdown of compact binaries including neutron
stars (\BNS and \NSBH systems). One should
include all \GW observations of compact binaries in order to extract the maximum
amount of information from the observations. In fact, as discussed above, we are
unable to distinguish models which vary the maximum allowed neutron star mass
since we ignore these events here. In this study we ignored these events since
the predicted detection rates for \BBH mergers dominated those of
other compact binary mergers. The \BBH mass distribution also spans
a large range of masses, with structure encoding information about binary
evolution. Ignoring other families of compact binaries also allowed us to avoid
ambiguities in discerning the family of the source (\BNS, \NSBH or \BBH) due to
degeneracies which exist in measuring the mass ratio for these systems
\citep{Hannam:2013uu}, although this can be dealt with in the future
\citep{2013arXiv1302.5341F}.

All these considerations have to be carefully taken
into account in future studies. \change{%
However, our results indicate that the upcoming generation of advanced \GW
detectors will soon start putting non-trivial bounds on current and future
binary evolution models, and analyses like the one presented here will
provide an important basis to link theoretical models with \GW observations.
}

\section*{Acknowledgements}

The authors would like to thank Mark Hannam, Ilya Mandel and Chris Messenger for
useful discussions. SS would like to acknowledge support from the STFC, Cardiff
University and the University of Birmingham. FO has been supported by the STFC
grant ST/L000962/1. SF would like to acknowledge
the support of the Royal Society and STFC grant ST/L000962/1.

\bibliography{calculatingProbability}

\end{document}